\documentclass[%
 reprint,
nofootinbib,
 amsmath,amssymb,
 aps
floatfix,
]{revtex4-2}
\usepackage{tikz}

\usepackage{graphicx}
\usepackage{dcolumn}
\usepackage{bm}
\usepackage{physics}
\usepackage{color}
\usepackage{graphicx}

\usepackage{float} 

\usepackage{hyperref}
\usepackage{nameref}
\usepackage[bottom]{footmisc}
\usepackage[mathlines]{lineno}

\begin{document}

\preprint{APS/123-QED}

\title{Constraining Inflationary Particle Production with CMB Polarization} 

\author{Luca H.~Abu El-Haj}
\email{lha2126@columbia.edu}
\affiliation{Department of Physics, Columbia University, NY 10027, USA}

\author{Oliver~H.\,E.~Philcox}

\affiliation{Leinweber Institute for Theoretical Physics at Stanford, 382 Via Pueblo, Stanford, CA 94305, USA}
\affiliation{Kavli Institute for Particle Astrophysics and Cosmology, 382 Via Pueblo, Stanford, CA 94305, USA}
\affiliation{Simons Society of Fellows, Simons Foundation, New York, NY 10010, USA}
\affiliation{Department of Physics, Columbia University, NY 10027, USA}

\author{J.~Colin Hill}

\affiliation{Department of Physics, Columbia University, New York, NY 10027, USA}

\begin{abstract}
\noindent  Following \citet{Philcox:2024jpd},
we investigate a scenario with a massive partner to the inflaton ($\mathcal{O}(100)$ times the inflationary Hubble scale), in which particles are produced during a narrow time period, leaving characteristic hot- or cold-spots in the cosmic microwave background (CMB).  Using tools developed for thermal Sunyaev-Zel'dovich cluster-finding, we search component-separated \emph{Planck} PR4 $E$-mode maps for these hotspots, and compare to analogous results in $T$. Our analysis pipeline is validated on simulated observations and gives unbiased constraints for sufficiently large and bright hotspots.  At \emph{Planck} sensitivities, the temperature data are more sensitive to small hotspots, but for sufficiently large hotspots the polarization data are more sensitive. We improve upon earlier work by building a full Poissonian likelihood for the hotspot abundance.  We find no strong evidence for primordial hotspots and thereby place novel bounds on the marginal and relevant couplings between the inflaton and massive scalars during inflation, probing physics at energies many orders of magnitude above any feasible terrestrial collider. The bounds derived from our new likelihood improve upon those of~\cite{Philcox:2024jpd} by more than an order of magnitude for sufficiently light particles ($M_0\lesssim100H_I$). We also forecast the inferred bounds on inflationary physics for a search using Atacama Cosmology Telescope (ACT) data, and from an optimistic cosmic-variance-limited experiment (CV), for which $E$-mode data provide stronger constraints than $T$ on nearly all scales. ACT should improve on the \emph{Planck} constraints by $\gtrsim10\%$, nearing the CV limit allowed by its sky coverage. Finally, we compare the constraining power of localized searches to that of a power spectrum analysis, and demonstrate that for sufficiently few produced particles the localized search performed herein is dominant.
\end{abstract}

\maketitle

\section{\label{sec:level1}Introduction}

\noindent A critical feature of inflation is that it generates fluctuations in the temperature ($T$) and $E$-mode polarization of the cosmic microwave background (CMB) in correlated ways, differing only by their respective transfer functions. Inflationary physics can thus be constrained by both observations of the CMB intensity and polarization, which crucially differ in their noise and systematic properties. In this work, we explore this in the context of a specific inflationary scenario involving an extremely massive scalar coupled to the inflaton.

Given the enormous number of microphysical inflationary scenarios, a natural way to constrain the physics of the early universe is to consider fairly generic properties of these models and to explore their phenomenological implications in the context of contemporary and future cosmological surveys. The existence of many fields during inflation, even extremely massive ones, 
is a common prediction of UV-complete physical models~\cite{baumann2012tasilecturesinflation,berglund2009multifieldinflationstringtheory}; the inflationary production of such particles thus provides a powerful phenomenological probe of many multi-field models. In practice, this production can be sourced by interactions between the inflaton and the massive field, which can induce a time-dependent effective mass for the latter. If this mass drops rapidly, particle production can be favorable~\cite{Kim:2021ida,Flauger_2017,Mirbabayi_2015,Kofman_2004}. Moreover, if the minimum effective mass is sufficiently large when particles are produced, the particles can measurably modify the local gravitational potential, or equivalently induce a curvature perturbation~\cite{Maldacena_2015}, which in turn evolves to produce localized hot- or cold-spots in the CMB\footnote{Following \cite{Philcox:2024jpd}, we refer to both hotspots and coldspots as ``hotspots'' for concision. As we will discuss in Sec.~\ref{sec:theory}, given the complicated profile structure for $E$-modes there is some ambiguity about whether particle production anisotropy is ``hot'' or ``cold''.}. The observational consequences of the production of massive particles have been studied in CMB temperature data~\cite{M_nchmeyer_2019,Kim:2021ida,Kim:2023wuk,Philcox:2024jpd}, but have not been extensively explored in polarization. 

If these hotspots exist, how can we best search for them in observational data? In the context of axion monodromy inflation, \citet{M_nchmeyer_2019} demonstrate that a matched-filter estimator approach for local profile-finding performs better in constraining inflationary particle production than $N$-point correlation function analyses in the large-$N$ limit, so long as particle production is rare. In this work, we follow a similar path, as outlined in~\cite{Philcox:2024jpd}, applying a matched-filter estimator to the \textsc{sevem} and \textsc{smica} component-separated $E$-mode maps built from the \emph{Planck} PR4 data~\cite{Planck:2020olo,Carron_2022,Philcox:2025wts}. We exploit the fact that, while it is uncommon to employ localized searches in polarization, we can form scalar $E$-mode maps analogous to conventional $T$ fields, so we may take advantage of data analysis parallels with searches for galaxy clusters in CMB maps via the thermal Sunyaev-Zel'dovich (tSZ) effect (the inverse-Compton scattering of CMB photons off hot electrons in the intracluster medium~\cite{1969Ap&SS...4..301Z}). In both the tSZ and the primordial hotspot production case, there is a well-defined angular profile of the signal of interest, characterized by a few free parameters. We compute these angular profiles for polarization and then construct a matched filter to search for them. 

Polarization provides an attractive probe with which to constrain inflationary physics, for a wide variety of reasons. Most obviously, polarization gives a (partially) statistically independent signal to temperature, from which we can glean new information and cross-check temperature results. Strikingly, hotspot profiles for temperature and polarization are very phenomenologically distinct, due to the  differences in their respective transfer functions. For this reason, coupled with the different properties of noise in $T$ and $E$, we will later find that, for 
future experiments (and existing ground-based CMB experiments), 
polarization data will provide better constraints on inflationary hotspots than temperature on nearly all scales. Although this may be model-specific,\footnote{However, we expect it is generically true for such particle production signatures, given works such as \cite{Galli_2014}, which demonstrate that CMB polarization constrains cosmological parameters better than temperature in the cosmic-variance limit.} it emphasizes the power of a polarization search, and makes clear that, as we enter a generation of CMB experiments where polarization sensitivity is comparable to temperature~\cite{ACT:2025xdm,ACT:2025tim,SPT-3G:2025bzu,SimonsObservatory:2018koc,SimonsObservatory:2025wwn}, localized polarization searches are powerful. Even beyond future interest, we show from simulations that our \emph{Planck} polarization search performs better on large angular scales 
than that of temperature and actually exceeds the constraining power of a temperature-only cosmic-variance-limited search on these scales. Additionally, foreground contamination is much less significant in polarization than in temperature, particularly on small scales (noting that tSZ clusters, which can cause false detections in temperature, are unpolarized to leading order). Polarization comes with an additional built-in cross-check: knowing that noise impacts $E$- and $B$-mode polarization similarly, and foregrounds can produce both, we could explicitly check for a strong hotspot candidate that it \emph{did not} appear in a $B$-mode map. While we do not find any sufficiently strong candidates in this work to motivate such a test, this could be very powerful in the event of a future detection.

Our matched-filter analysis would, in principle, produce a catalog of primordial hotspots, although in practice we do not find any robust detections.  Nevertheless, the measured hotspot abundance can then be compared to theoretical predictions to derive constraints on the underlying parameters of the model, in direct analogy to cluster abundance cosmology~\cite[e.g.,][]{2001ApJ...560L.111H}.  Thus, we construct a Poissonian likelihood for the hotspot abundance, which we use to place constraints on this inflationary particle production mechanism. The introduction of this new likelihood allows us to derive significantly stronger bounds than found in previous works studying this scenario, and provides a systematically useful new tool for constraining similar mechanisms.

The remainder of this work is organized as follows. In Sec.~\ref{sec:theory}, we discuss particle production hotspots and their phenomenology. In Sec.~\ref{sec:meth}, we explain our pipeline for the polarization search. We then verify our pipeline on simulations in Sec.~\ref{sec:sims} and apply it to \emph{Planck} \textsc{sevem} and \textsc{smica} component-separated maps in Sec.~\ref{sec:Planck}, comparing to the search using temperature data performed in \cite{Philcox:2024jpd}.  In Sec.~\ref{sec:forecasts}, we carefully explore the consequences of polarization versus temperature data in localized hotspot searches in general and robustly forecast the detection sensitivities of both probes for current and future CMB experiments, and in Sec.~\ref{sec:bounds}, we use these sensitivities to derive constraints on the underlying physical parameters of the theory via a Poissonian likelihood for the hotspot abundance. In Sec.~\ref{sec:powerspec}, we explore the constraining power of a power spectrum analysis, and compare it to our profile-finding method for both temperature and polarization, before concluding in Sec.~\ref{sec:summary}. Throughout, we adopt a flat $\Lambda$CDM cosmology based on~\cite{Tristram:2023haj}: $\{H_0=67.32~\text{km}/\text{s}/~\text{Mpc}, \omega_\text{b}=0.022383, \omega_{\text{cdm}}=0.12011, \tau_{\text{reio}}=0.0543, \sum m_\nu=0.06~\text{eV}, A_s=2.1006\times 10^{-9}, n_s=0.96605\}$. We use the $(-,+,+,+)$ metric signature convention.

\section{Theoretical Background}
\label{sec:theory}

\subsection{Particle Production}
\noindent Let us consider an extremely massive field $\sigma $ whose effective mass momentarily passes through a minimum at some inflaton field value $\varphi_*$, yielding the expansion:
\begin{align}
    M_\sigma(\varphi)=M_0+\dv[2]{M_\sigma}{\varphi}\bigg\rvert_{\varphi_*} \frac{1}{2}(\varphi-\varphi_*)^2+\mathcal{O}((\varphi-\varphi_*)^3) \,.
\end{align}
The validity and interest of this approximation can equivalently be thought of as the inclusion of only relevant and marginally relevant terms in the effective mass term of the Lagrangian for the $\sigma$ field, which is given by 
\begin{align}
    \mathcal{L}\supset -\frac{1}{2}(\partial\sigma)^2-\frac{1}{2} \left(M_0^2+(g\varphi-\mu)^2\right)\sigma^2 \,,
    \label{eqn:lagrangian}
\end{align}
where we have defined coupling constants $g^2\equiv M_0\dv[2]{M_\sigma}{\varphi}\rvert_{\varphi_*}$ and
$\mu\equiv g\varphi_*$.
Employing the slow-roll expansion, we may write $\varphi-\varphi_*\simeq\dot{\varphi_0}(t-t_*)$, with the slow-rolling velocity given by $\dot{\varphi}_0 \simeq (57H_I)^2$ from the large-scale CMB anisotropy normalization, where $H_I$ is the inflationary Hubble scale. Under this approximation, the effective time-dependent squared mass is given by: 
\begin{align}
    M^2_\sigma\simeq M_0^2 + g^2\dot{\varphi_0}^2(t-t_*)^2 \,.
\end{align}
In conformal time, defined by $\eta \equiv \int \frac{\dd t}{a}$ where $a$ is the scale factor (during standard quasi-de Sitter inflation $a\sim e^{H_It}$), we may simplify by noting that $(t-t_*)=-(1/H_I)\ln(\eta/\eta_*)$, where $\eta_*$ denotes the conformal time at production\footnote{While formally this is the relevant definition, more intuitively $\eta_*$ defines a characteristic size for the hotspot --- see Figure~\ref{fig:widefigure}.}:  
\begin{align}
   M_\sigma^2\simeq M_0^2 + \frac{g^2\dot{\varphi}_0^2}{H_I^2}\ln^2{\abs{\frac{\eta}{\eta_*}}} \,.
\end{align}
For a simple estimate, if we take $M_0=\mathcal{O}(100H_I)$ and $g=\mathcal{O}(10)$, consistent with the values explored in this work, it is clear that the second term dominates, and in particular as $\eta\rightarrow\eta_*$ the time-dependent term vanishes so there is a significant decrease in the effective mass, thus allowing $\sigma$ particle production. 

\begin{figure}[t]
  \centering

\includegraphics[width=0.5\textwidth]{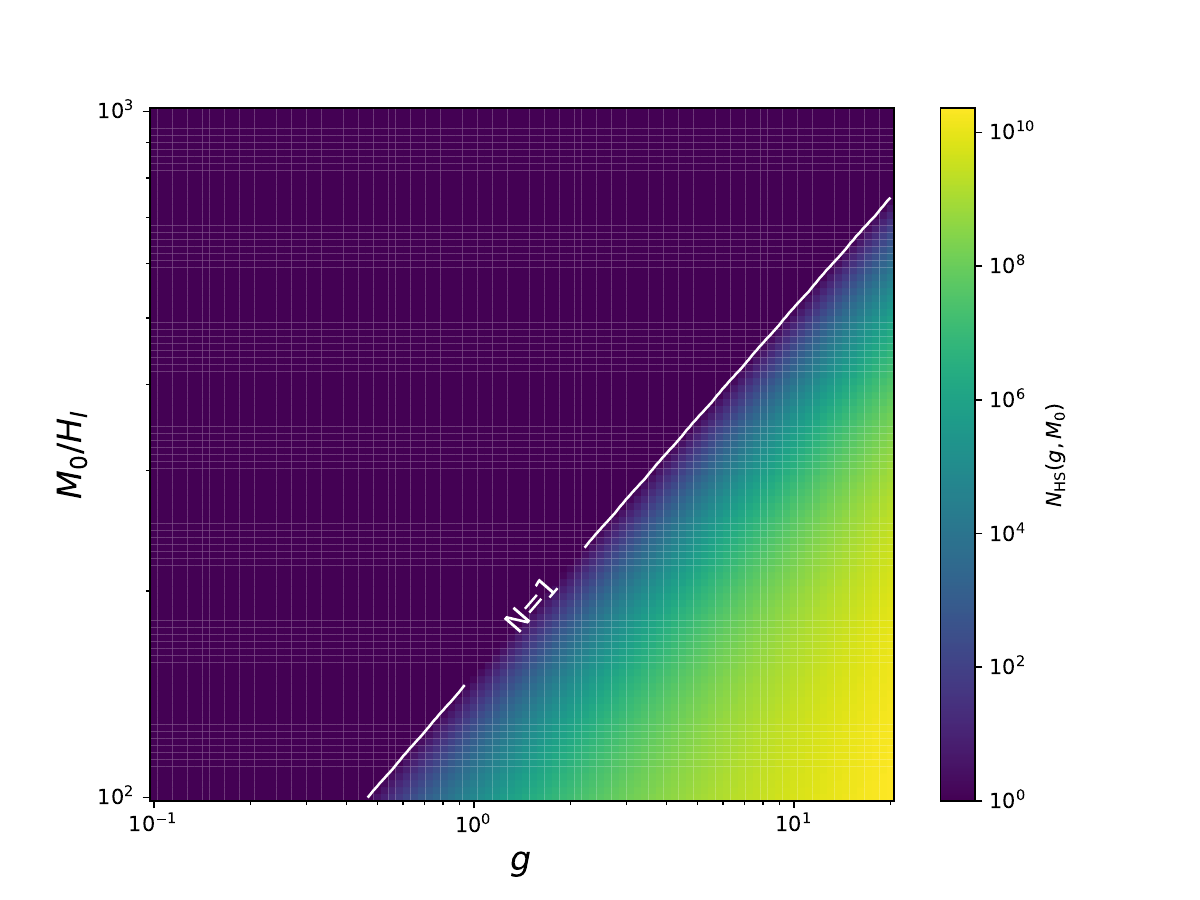}
  \caption{The mean number of particle-production hotspots $N_{{\mathrm{HS}}}(g,M_0)$ { as a function of the coupling $g$ and minimal mass $M_0$}, where we take characteristic values for $\eta_*$ and $\Delta\eta$ equal to 100 Mpc to give a sense of the nontrivial dependence of particle production on the physical parameters in the model. We highlight the $N_{\rm HS} = 1$ contour, which we will find is the parameter space in which our single-hotspot sensitivities derived in Sec.~\ref{sec:forecasts} and the population-level bounds in Sec. \ref{sec:bounds} should be equivalent}.
  \label{fig:N_of_g}
\end{figure}

The particle production associated with this Lagrangian can be computed using standard Bogoliubov techniques. A pedagogical review of the calculation can be found in~\cite{Kim:2021ida}. The approximate number of hotspots produced in a shell of thickness $\Delta\eta$ around the surface of last scattering is given by~\cite{Kim:2021ida,Kim:2023wuk} 
\begin{align}\label{eqn:NHS}
    N_{{\mathrm{HS}}}\simeq 4\times 10^{8}\times g^{\frac{3}{2}}\left(\frac{\Delta\eta}{100 \,\,\text{Mpc}}\right) \left(\frac{100 \,\,\text{Mpc}}{\eta_*}\right)^3\\
    \times e^{-\pi(M_0^2-2H_I^2)/(g\abs{\dot{\varphi_0}})}\nonumber \,.
\end{align}
The number of hotspots is clearly dependent on the mass $M_0$ of the particle.  For $M_0 \gg H_I$, one can assume that the produced particles become non-relativistic very quickly, and as such induce localized features on the CMB.  In Figure~\ref{fig:N_of_g}, we plot $N_{{\mathrm{HS}}}$ as a function of $g$ and $M_0/H_I$, for representative values $\eta_* = 100~\text{Mpc} = \Delta \eta$. As we discuss in Sec.~\ref{sec:powerspec}, for a large number of hotspots (shown in the yellow region of parameter space in Figure~\ref{fig:N_of_g}, for example) a power spectrum analysis becomes more optimal than a localized search.  Our profile-finding search is optimal in the upper {left region} of parameter space shown in Figure~\ref{fig:N_of_g}{, where fewer particles are produced.}

\subsection{CMB $E$-mode Particle-Production Hotspots}
\noindent The hotspots induced by massive particle production can be thought of intuitively as follows. The production of a very massive particle induces a local gravitational potential, which, in turn, causes a non-zero vacuum expectation value for the curvature perturbation, which propagates to the observable CMB anisotropies. Equivalently, one can consider the Lagrangian in Equation~\eqref{eqn:lagrangian} as containing an interaction term between $\varphi$ and $\sigma$. As such, $\sigma$ exerts a force on the inflaton field, slowing it down and locally causing inflation to end later. This leaves overdensities, which, after CMB decoupling, induce hot or cold spots depending on the CMB transfer function. These hotspot profiles for temperature have been computed and explored extensively in~\cite{Philcox:2024jpd,Kim:2021ida,Kim:2023wuk}; here we present the analogous computation for $E$-mode polarization. 

Adopting a similar approach to previous works, one can analytically compute the position-space profiles for these hotspots. The curvature perturbation associated with a massive particle is computed in~\cite{Kim:2021ida} and reproduced for convenience in Appendix~\ref{Appendix:appendix A}; this yields 
\begin{align}
    \langle\zeta_{\mathrm{{\mathrm{HS}}}}(\mathbf{k})\rangle= e^{-i\vb{k}\vdot \vb{x_{\mathrm{{\mathrm{HS}}}}}}\frac{gH_I^2}{\dot{\varphi_0}} \frac{\text{Si}(k\eta_*)-\sin{k\eta_*}}{k^3} \,.
\end{align}
Here, $\text{Si}(x)\equiv\int_0^x\frac{\sin{t} }{t}\dd t$ and we also define $f(x)\equiv\frac{gH_I^2}{\dot\varphi_0} \left(\text{Si}(x)-\sin{x} \right)$. To transform from curvature to CMB observables, we must take into account the $E$-mode transfer function $\mathcal{T}^E_\ell(k)$, which is dominated by Doppler contributions. 
The $E$-mode anisotropy induced by a curvature perturbation from a Fourier mode of wavevector $\bf{k}$ is~\cite{2020moco.book.....D,Weinberg:2008zzc} 
\begin{align}
    \delta E(\vb{k},\hat{\bf{n}})=\sum_{\ell=2}^{\infty}i^\ell\mathcal{P}_\ell(\hat{\bf{n}}\vdot \vb{k})\mathcal{T}^E_\ell(k) \langle \zeta_{\mathrm{{\mathrm{HS}}}}(\mathbf{k})\rangle \,,
\end{align}
where $\mathcal{P}_\ell(x)$ are the Legendre polynomials. Summing over Fourier modes (following~\cite{Kim:2023wuk}) leaves us with
\begin{align}
\label{eq:profile}
\delta E(\hat{\bf{n}},\hat{\bf{n}}_{{\mathrm{HS}}},\eta_*,\eta_{{\mathrm{HS}}})&=\frac{1}{2\pi^2}\sum_{\ell=2}^\infty (2\ell+1)\mathcal{P}_\ell(\vb{\hat{\bf{n}}}\vdot\vb{\hat{\bf{n}}_{{\mathrm{HS}}}})\\&\times\int\frac{\dd k}{k}f(k\eta_*) j_\ell(k\chi_{{\mathrm{HS}}})\mathcal{T}^E_\ell(k)\nonumber \,,
\end{align}
where $j_\ell(x)$ are the spherical Bessel functions. Here we have approximated the position of the hotspot as $\vb{x}_{{\mathrm{HS}}}-\vb{x}_0 \simeq (\chi_{{\mathrm{HS}}}, \hat{\vb{n}}_{{\mathrm{HS}}})$,~{where $\chi_{{\mathrm{HS}}}=\eta_0-\eta_{{\mathrm{HS}}}$ is the comoving distance to the hotspot and $\eta_{{\mathrm{HS}}}$ and $\eta_0$ are respectively the conformal time at the hotspot production epoch and today.} We assume that $\eta_{{\mathrm{HS}}}$ is close to the {conformal time} at last scattering $\eta_\text{rec}$, such that the massive particles leave detectable impacts on the observed CMB.  A sketch of the geometry relevant to the problem is shown in Figure~\ref{fig:geometry}. We compute the transfer function using $\textsc{camb}$\footnote{\url{https://camb.info}} 
and cut off the sum above $\mathrm{\ell}_{\max}=3500$, which is sufficient to extract all information in contemporary and near-future CMB experiments. Note critically the linearity of the profile in $g$, which is the basis of our analysis pipeline.

\begin{figure}[t]
    \centering
\begin{tikzpicture}[scale=2]

\def\Rrec{1.3}
\def\deta{0.3}
\def\Rin{\Rrec - \deta}
\def\Rout{\Rrec + \deta}

\def\rHS{1.38}
\def\angleHS{30}

\coordinate (x0) at (0,0);
\node[below left] at (x0) {$\vb{x}_0$};

\fill[even odd rule, gray!25] (x0) circle(\Rout) (x0) circle(\Rin);

\draw[line width=1pt] (x0) circle(\Rrec);

\node (chirecLabel) at (0,-\Rrec-0.12) {$\chi_{\mathrm{rec}}$};

\draw[->][thick] (x0) -- (chirecLabel.north);

\draw[->][thick] (\Rrec,0.10) -- (\Rout,0.10)
    node[midway, above] {$\eta_*$};

\coordinate (xHS) at (\angleHS:\rHS);
\fill[red] (xHS) circle (0.035);
\node[right] at (xHS) {$\vb{x}_{\mathrm{HS}}$};

\draw[->][thick] (x0) -- (xHS)
    node[midway,left] {$\chi_{\mathrm{HS}}$};

\end{tikzpicture}
    \caption{Sketch of the relevant geometry, where $\vb{x}_0$ represents our position today, $\vb{x}_{\mathrm{HS}}$ is the hotspot position, $\chi_{\mathrm{rec}}=\eta_0-\eta_{\mathrm{rec}}$ is the comoving distance to the surface of last scattering, $\chi_{\mathrm{HS}}=\eta_0-\eta_{\mathrm{HS}}$ is the comoving distance to the hotspot, and $\eta_*$ is the comoving horizon size at particle production. Note that to causally impact the CMB the comoving distance between the hotspot and the surface of last scattering must be less than the horizon size (indicated by the gray band) at particle production, apart from small ISW contributions in $T$ for hotspots with $\eta_{\rm HS} - \eta_{\rm rec} > \eta_*$.}  
    \label{fig:geometry}
\end{figure}

\begin{figure*}[t!]
  \centering  
  \begin{minipage}{0.48\textwidth}
    \centering
    \includegraphics[width=\linewidth]{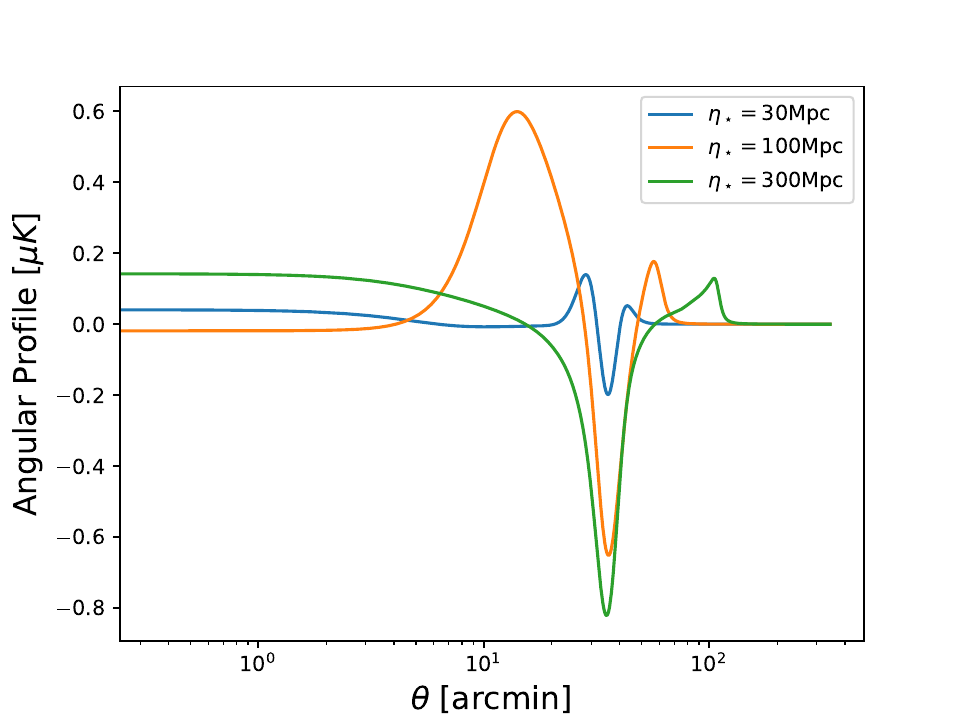}
    \vspace{2pt}
    
    \textbf{A})~\small {$E$-mode profile variation with respect to $\eta_*$, with $\eta_{{\mathrm{HS}}} = \eta_{\rm rec}$.}
  \end{minipage}
  \hfill
  \begin{minipage}{0.48\textwidth}
    \centering
    \includegraphics[width=\linewidth]{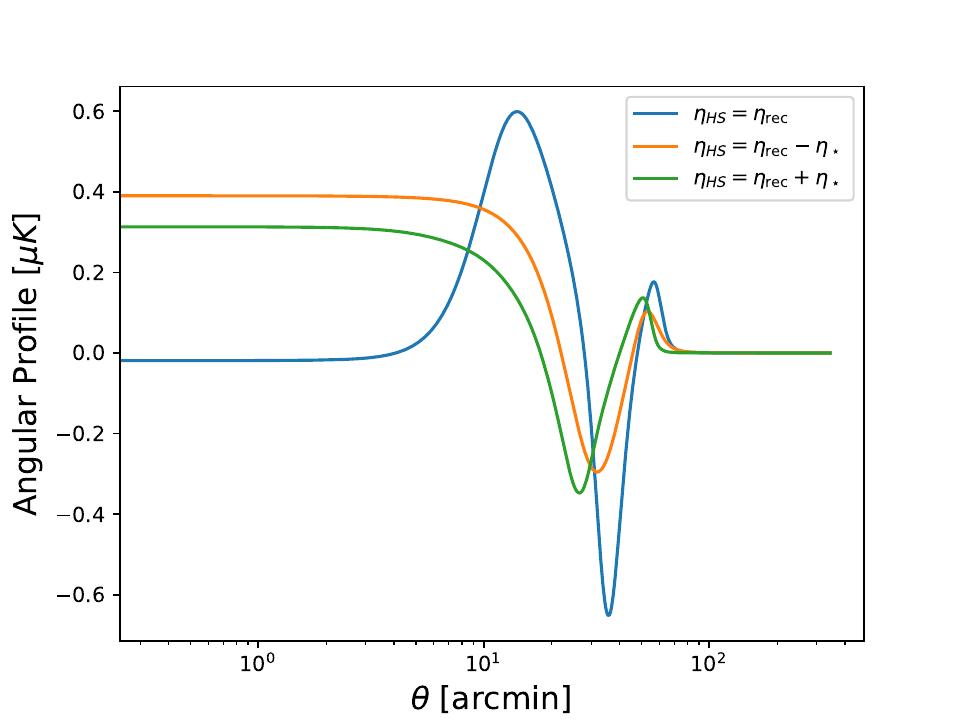}
    \vspace{2pt}
    
    \textbf{B})~\small $E$-mode profile variation with respect to $\eta_{{\mathrm{HS}}}$, with $\eta_* = 100$~Mpc.
  \end{minipage}
  \hfill
  \begin{minipage}{0.48\textwidth}
    \centering
    \includegraphics[width=\linewidth]{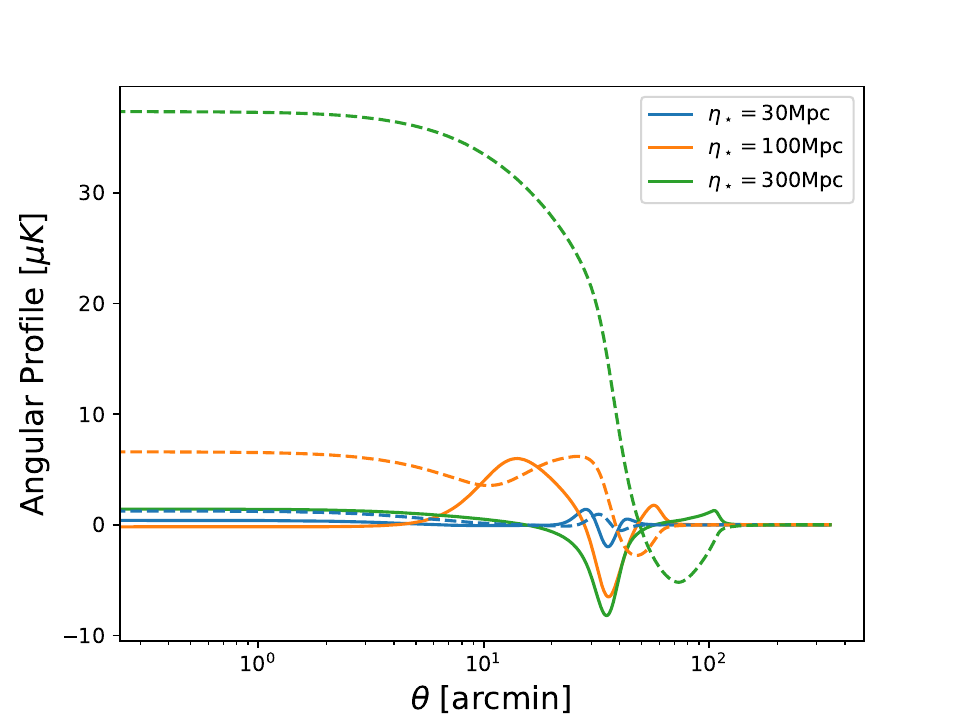}
    \vspace{2pt}
    
    \textbf{C)}~\small 
    Comparison of $T$ (dashed) and $E$ (solid) profiles, {with the latter multiplied by 10 for visual clarity}.
    
  \end{minipage}
  \caption{Temperature and polarization hotspot templates for $g=1$, demonstrating the variability of the polarization hotspot profiles with respect to both $\eta_*$ (encoding the hotspot formation time during inflation) and $\eta_{{\mathrm{HS}}}$ (parametrizing the hotspot distance). (\textbf{A}): We fix $\eta_{{\mathrm{HS}}}=\eta_{\text{rec}}$ and show profiles for $\eta_*=30,100,300\,\,\text{Mpc}$ in blue, orange, and green, respectively. (\textbf{B}): We fix $\eta_*=100~\text{Mpc}$ and show the variability of the profile with respect to $\eta_{{\mathrm{HS}}}$. Each hotspot has a significant ``cold ring" at $\theta \simeq 0.5$--$0.6^\circ$ from the center. (\textbf{C}): We compare the profiles for temperature (dashed) and polarization (solid). We multiply the polarization profiles by a factor of 10 for visual clarity. Notice that the temperature amplitude is characteristically around an order of magnitude larger than that in polarization.}
  \label{fig:widefigure}
\end{figure*}

\begin{figure*}[t]
  \centering

  \begin{minipage}{0.4\textwidth}
    \centering
    \includegraphics[width=\textwidth]{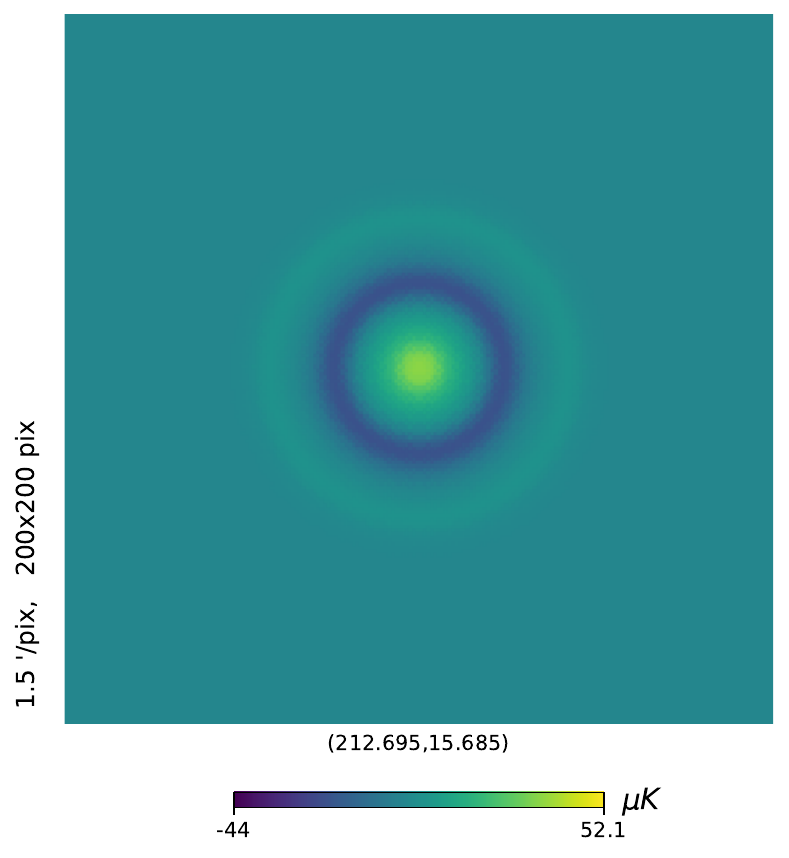}
    
    \textbf{A})~\small{Simulated $E$-mode hotspot signal.}
  \end{minipage}
  \hspace{0.5in}
  \begin{minipage}{0.4\textwidth}
    \centering
    \includegraphics[width=\textwidth]{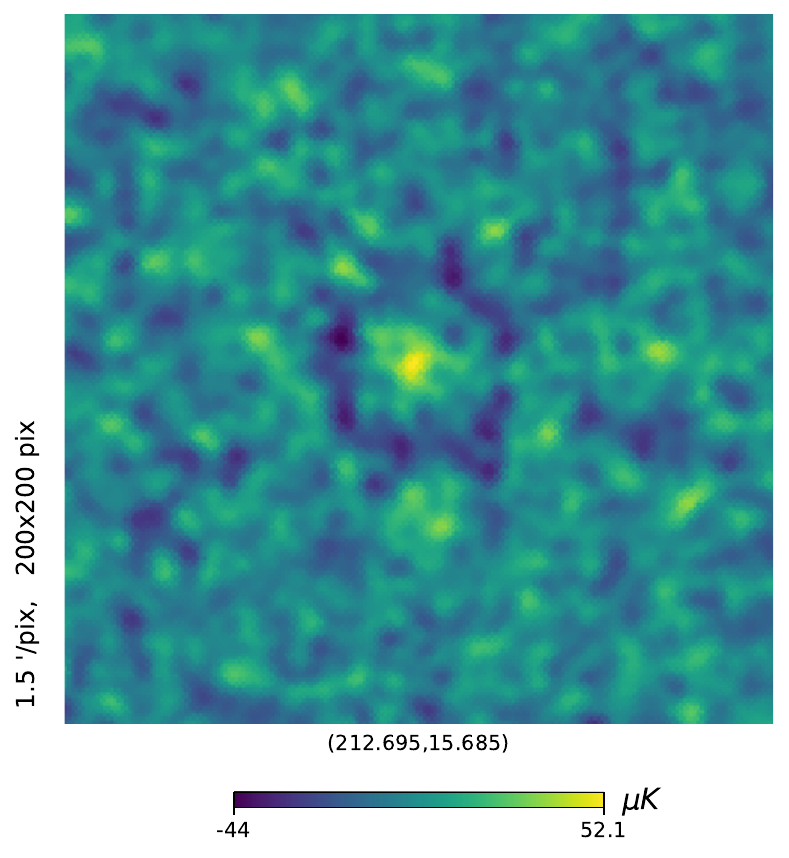}
    
    \textbf{B)}~\small{Simulated $E$-mode hotspot signal combined with CMB.}
  \end{minipage}

  \vspace{1em}

  \begin{minipage}{0.4\textwidth}
    \centering
\includegraphics[width=\textwidth]{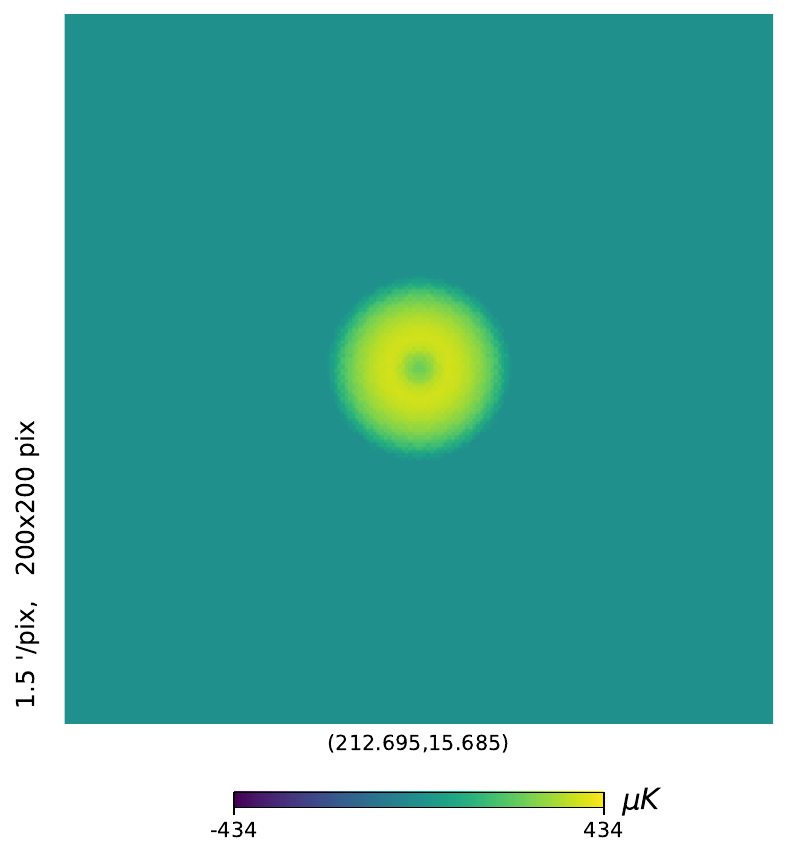}
    
    \textbf{C)}~\small{Simulated temperature hotspot signal.}
  \end{minipage}
  \hspace{0.5in}
  \begin{minipage}{0.4\textwidth}
    \centering
    \includegraphics[width=\textwidth]{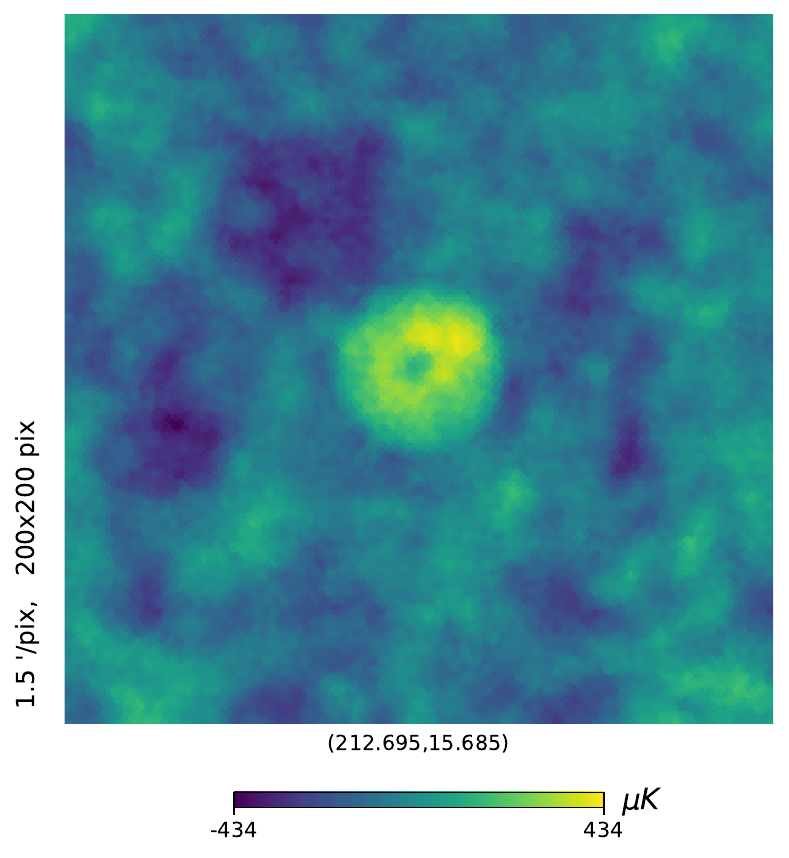}
     
    \textbf{D)}~\small{Simulated temperature hotspot signal combined with CMB.}
  \end{minipage}

  \caption{
    Exemplar simulated hotspots 
    with $g=30$, $\eta_*=129\ \mathrm{Mpc}$, and $\eta_{\mathrm{HS}}=\eta_{\mathrm{rec}}$.
    Left panels show signal-only maps; right panels show the same signals added
    to \textsc{npipe} simulations. The $T$ and $E$ hotspots are centered on the same point, and are shown in $5^\circ\times5^\circ$ maps with 1.5 arcminute pixels in the Gnomonic projection. 
  }
  \label{fig:simexamps}
\end{figure*}

We provide exemplar $E$-mode hotspot profiles in Figure~\ref{fig:widefigure}, and examples of simulated hotspot profiles on the sky in Figure~\ref{fig:simexamps}. The $E$-mode profiles have a number of salient features. Beyond just a central amplitude, the profiles have a characteristic trough around 36 arcmin (0.01 radians). This trough is $\eta_{{\mathrm{HS}}}$-dependent; for these calculations, we fix $\eta_{{\mathrm{HS}}}=\eta_{\text{rec}}$, though other values will be discussed below. There is also a characteristic, though significantly less dramatic, peak at a small angular distance past the trough. Notice also that these are phenomenologically quite different from the temperature hotspot profiles (also shown in Figure~\ref{fig:widefigure}), where the characteristic feature is the central temperature and the profile cuts off around the characteristic scale $\theta_*=\sqrt{4\pi}\eta_*/\eta_0$.

\begin{figure*}[t]
  \centering
  \includegraphics[width=0.48\textwidth]{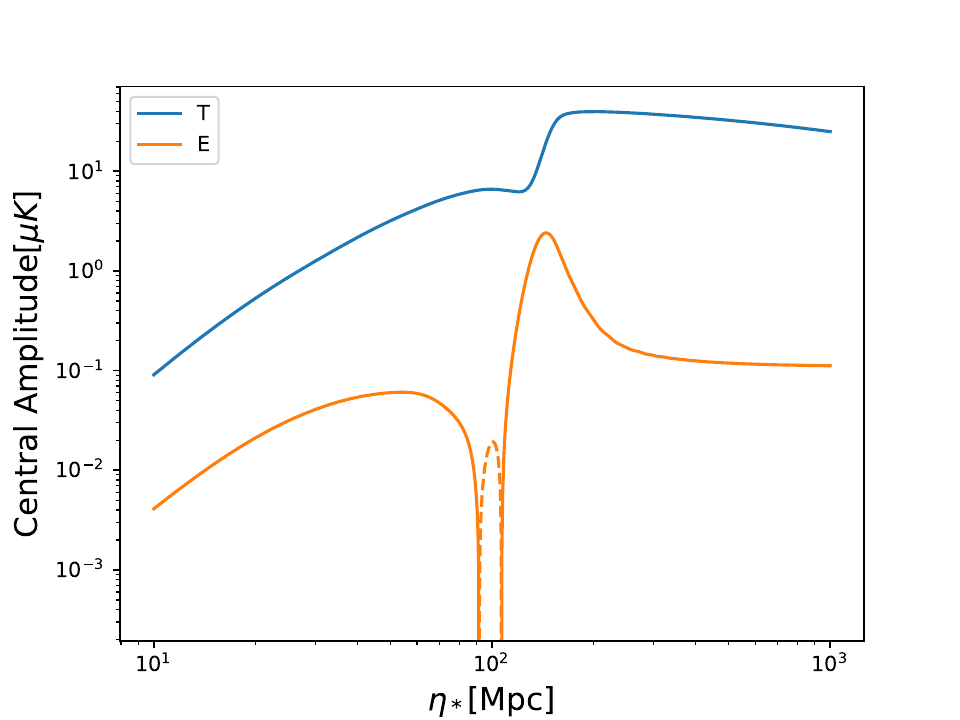}
  \hfill
  \includegraphics[width=0.48\textwidth]{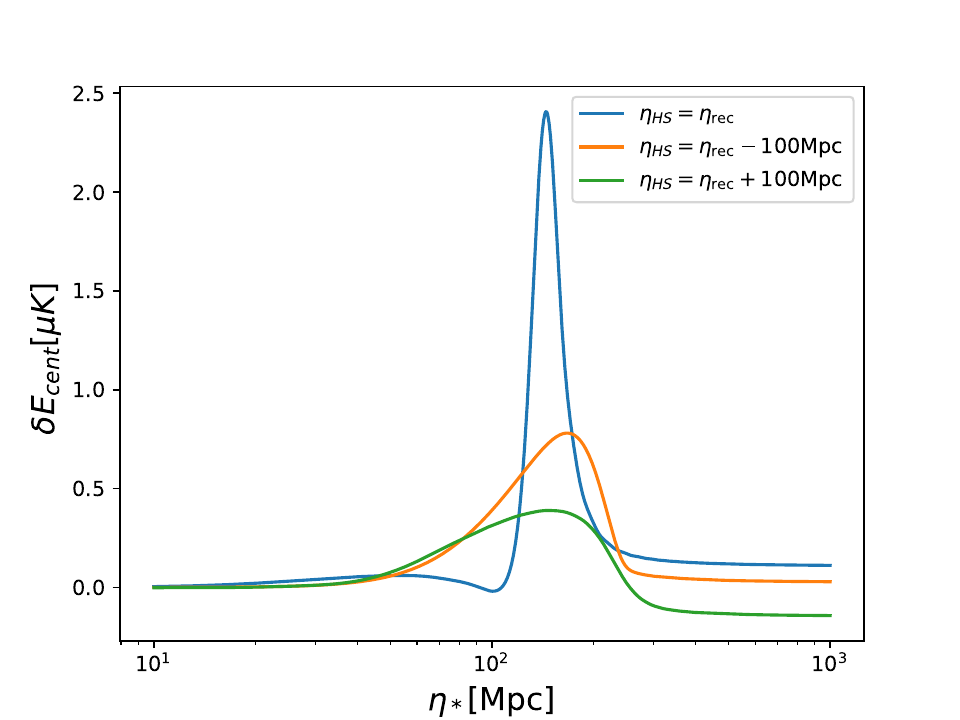}
  \caption{Central amplitude of the hotspots. \textbf{(Left)}: We show a comparison between the central amplitudes of hotspot profiles in temperature (blue) and in $E$-mode polarization (orange), where we have fixed $\eta_{{\mathrm{HS}}}=\eta_{\text{rec}}$. Negative values are plotted with dashed curves. \textbf{(Right)}: We show the central amplitude for polarization computed for various $\eta_{{\mathrm{HS}}}$ values, where we have chosen $100$~Mpc as a characteristic scale for the variation of $\eta_{\mathrm{HS}}$ away from $\eta_{\mathrm{rec}}$, while preserving causality. Note that the central amplitude of the anisotropy is strongest for hotspots located close to the surface of last scattering ($\eta_{{\mathrm{HS}}}=\eta_{\rm rec}$). 
  }
  \label{fig:widefigure_centemps}
\end{figure*}
For a better sense of the phenomenology, we also present in Figure~\ref{fig:widefigure_centemps} the central amplitude of the $E$-mode hotspot profile, which is computed by taking $\hat{\bf{n}}\vdot \hat{\bf{n}}_{{\mathrm{HS}}}\rightarrow1$, which means that $\mathcal{P}_\ell\rightarrow1$. Thus,
\begin{align}
    \delta E_{\text{cent}} = \frac{1}{2\pi^2} \sum_{\ell=2}^\infty (2\ell+1)\int\frac{\dd k}{k}f(k\eta_*) j_\ell(k\chi_{{\mathrm{HS}}})\mathcal{T}^E_\ell(k) \,.
\end{align}
We note two phenomenological features in Figure~\ref{fig:widefigure_centemps}. First, we find a strongly peaked behavior for the polarization central amplitude {at $\eta_*\approx 150\,\mathrm{Mpc}$, corresponding to modes that re-entered the horizon at roughly half the conformal time of last scattering.  This value of $\eta_*$ is approximately equal to the sound horizon at last scattering, $r_s^{\rm rec}$.} Additionally, we see in the right panel of Fig.~\ref{fig:widefigure_centemps} that the central amplitude is peaked at $\eta_{\rm{HS}}=\eta_{\rm{rec}}$, near where the CMB visibility function peaks.

Also, we find the temperature anisotropy to be characteristically $\sim10\times$ larger than the polarization, similar to that of the ratio between polarization and temperature from a standard single-field inflationary scenario. Considering the variation with respect to $\eta_{\rm HS}$, one should also observe that the central amplitude is sharply peaked around $\eta_{\mathrm{HS}}=\eta_{\mathrm{rec}}$.  This arises from the $E$-mode polarization transfer function, which is sharply peaked around recombination.

\section{\label{sec:meth} Analysis Methodology}
\noindent To search for inflationary hotspots in $E$-mode data, we use a methodology similar to~\cite{Philcox:2024jpd}. By conservation of momentum in a homogeneous inflationary background, our particle production mechanism should produce pairs of particles separated by a distance $\alt \eta_*$~\cite{Kim:2021ida}. An optimal search would thus involve a pairwise profile, which is computationally intensive since (a) the pairwise profile breaks isotropy, and (b) we do not know the separation precisely, given its heavy dependence on the microphysics of the problem. We simplify this by performing a search for single hotspots, which is shown to be near-optimal by~\cite{Philcox:2024jpd,Kim:2023wuk}.

To constrain the coupling parameter $g$, we use a matched-filter analysis, which is implemented following the methodology of tSZ cluster searches~\cite{Planck:2015vgm}. We make minor modifications to the \textsc{szifi} code described in~\cite{Zubeldia:2022gva,Zubeldia_2023,Philcox:2024jpd}, enhancing the primordial non-Gaussianity functionality added in \citep{Philcox:2024jpd} allowing for the analysis of both $T$ and $E$ maps. This amounts to the introduction of a massive particle hotspot template in $E$, and the addition of functionality for the $E$-mode \emph{Planck} \textsc{sevem} and \textsc{smica} component-separated maps, using the \emph{Planck} \textsc{smica} polarization beam. 

Our analysis is performed on \textsc{sevem}~\cite{Planck:2020olo} and \textsc{smica}~\cite{Carron_2022,Philcox:2025wts} \emph{Planck} PR4 component-separated $E$-mode polarization maps produced using a spin-two harmonic transform from $Q/U$ to real space $E/B$-modes in \textsc{healpy}~\cite{Gorski_2005} with $N_{\rm side}=2048$. This transformation allows for a scalar treatment of real space $E$-mode maps. We then apply the  \emph{Planck} component-separated common mask to the $E$-mode real-space maps, which removes the Galactic plane, and we inpaint a polarized point source mask using a diffusive algorithm~\cite{mccarthy2024componentseparatedcibcleanedthermalsunyaevzeldovich}. Once the maps have been produced and masked, we split the full sky into 768 tiles, each $14.8^\circ\times14.8^\circ$. Each tile contains $1024^2$ pixels following the procedure of~\cite{Zubeldia:2022gva,Zubeldia_2023}. We then use a matched filter estimator for the coupling constant:
    \begin{align}
       \hat{g}(\boldsymbol{\theta})&=\sigma_g^2\int \frac{\dd ^2\vb{l}}{(2\pi)^2}\frac{t^*(\vb{l},\boldsymbol{\theta})d(\vb{l})}{C^{EE}(\vb{l})} \,,\\
        \sigma_g&=\left( \int \frac{\dd^2 \vb{l}}{(2\pi)^2}\frac{\abs{t(\vb{l,\vb{0}})}^2}{C^{EE}(\vb{l})} \right)^{-\frac{1}{2}} \,.
        \label{eq:MMF}
    \end{align}
Here, $t(\vb{l},{\boldsymbol \theta})$ is the Fourier-space hotspot profile centered at ${\boldsymbol \theta}$, and $d({\boldsymbol\theta})$ is the masked and inpainted component-separated $E$-mode map. By isotropy, we can fix the hotspot template at the origin. The power spectrum $C^{EE}(\vb{l})$, which includes both signal and noise, is directly computed from the data on each tile, via $C^{EE}(\vb{l})=\langle d(\vb{l}) d^*(\vb{l})\rangle$. The SNR of a hotspot candidate is defined by SNR $\equiv \hat{g}/\sigma_g$. Although we include transfer functions up to $\ell_{\rm max}=3500$ in the computation of our profiles, when computing $\hat{g}$ and $\sigma_g$ we only integrate in the range $\ell=[30,3000]$.     

We generate 100 position-space templates, using 10 logarithmically spaced values of $\eta_*\in[10,1000]~\text{Mpc}$ and similarly 10 linearly spaced values of $\chi_{{\mathrm{HS}}}\in[\chi_{\rm rec}-\eta_*,\chi_{\rm rec}+\eta_* ]$. We enforce a causality condition on the comoving distance by demanding $0\leq\chi_{{\mathrm{HS}}}\leq \eta_0$, where $\chi(\eta)=\eta_0-\eta$. Our lower bound is justified by noting that the $\emph{Planck}$ beam makes hotspots smaller than $\eta_* \approx 10~{\rm Mpc}$ undetectable. Additionally, we choose an upper bound of $1~\text{Gpc}$ based on the angular size of our tiles. Searching for larger hotspots would require going beyond the flat-space Fourier transform approximation, {and the number of such hotspots are cubically suppressed (in $\eta_*$) from Eq. \ref{eqn:NHS}}.

We compute position-space profiles using \textsc{camb}~\cite{2011ascl.soft02026L} transfer functions, across a logarithmically-spaced grid with $k\in[10^{-6},1]~h^{-1}\text{Mpc}$, using a Simpson integrator. We compute each profile on a $1024^2$ pixel grid, using a flat-space distance $\theta=\cos^{-1}({\hat{\bf{n}}\vdot \hat{\bf{n}}_{{\mathrm{HS}}}})$. We then convolve each profile with the \emph{Planck} \textsc{smica} beam. We compute each profile to a maximum $\theta_{\text{max}}=\text{max}[0.1~{\rm rad},\sqrt{4\pi}(\eta_*/\eta_0)]$. This avoids the edges of tiles, and also generally removes regions where the profile is less than 1\% of its peak~\cite{Philcox:2024jpd}.  While $T$ and $E$ have phenomenologically different profiles, the angular scale of the hotspots should be comparable (see Figure~\ref{fig:widefigure}). 

For the $E$-mode search, we restrict our analysis to component-separated data, rather than considering an analysis on the individual frequency maps using a multi-frequency matched filter (MMF). One reason for this is that to leading order there is no tSZ contamination in polarization.  Also, in general, there is significantly less small-scale foreground contamination in polarization than in temperature. Note that our use of the common mask excludes point sources in polarization with SNR~$\geq 5$.

To the output catalog we apply a mask that is frequently used for tSZ cluster searches, which removes the brightest $\simeq 40\%$ of the sky, as detailed in~\cite{Philcox:2024jpd,Planck:2015vgm}. This procedure leaves us with an output catalog of hotspot candidates, which we then further process by applying the \emph{Planck} HFI temperature point source mask~\cite{Planck:2020olo}, which excludes candidates within $10$ arcmin of $5\sigma$ point sources in temperature. We do this to avoid potential contamination associated with the slight polarization of a point source that is bright in temperature. In practice, as seen below in Sec.~\ref{sec:Planck}, we find no additional exclusions from this procedure.

\section{Simulation Results and Pipeline Validation}
\label{sec:sims}
\noindent Before jumping into the analysis on real \emph{Planck} maps, we verify our methodology on a single \textsc{npipe} simulation representing the full component-separated \emph{Planck} PR4 dataset. Unlike~\cite{Philcox:2024jpd}, we only consider simulations for our analysis using injected maps of single hotspots, \textit{i.e.}, we do not inject pairwise hotspot signals. This is justified given that~\cite{Philcox:2024jpd} found consistent recovery results between single and pairwise searches. 

We follow a simulation procedure similar to \cite{Philcox:2024jpd}, where we generate 300 $E$-mode hotspots with $\eta_{{\mathrm{HS}}}=\eta_{\rm rec}$ and consider ten logarithmically spaced values of $\eta_*\in[10,1000]~\text{Mpc}$, convolving all profiles with the \emph{Planck} \textsc{smica} beam.  We also demand all the injected hotspots to be more than $3^\circ$ apart, so as to avoid interference effects between two nearby hotspots. We then add this hotspot injection map to simulated \textsc{npipe} \textsc{sevem} $E$-mode polarization maps for $g\in[10,20,30,40,50]$, and run our analysis pipeline.  This requires around $20~\text{CPU}$-hours for each value of $g$, {where we have fixed $\eta_{\rm{HS}}=\eta_{\rm{rec}}$}.

\begin{figure*}[t]
  \centering
  \includegraphics[width=0.32\textwidth]{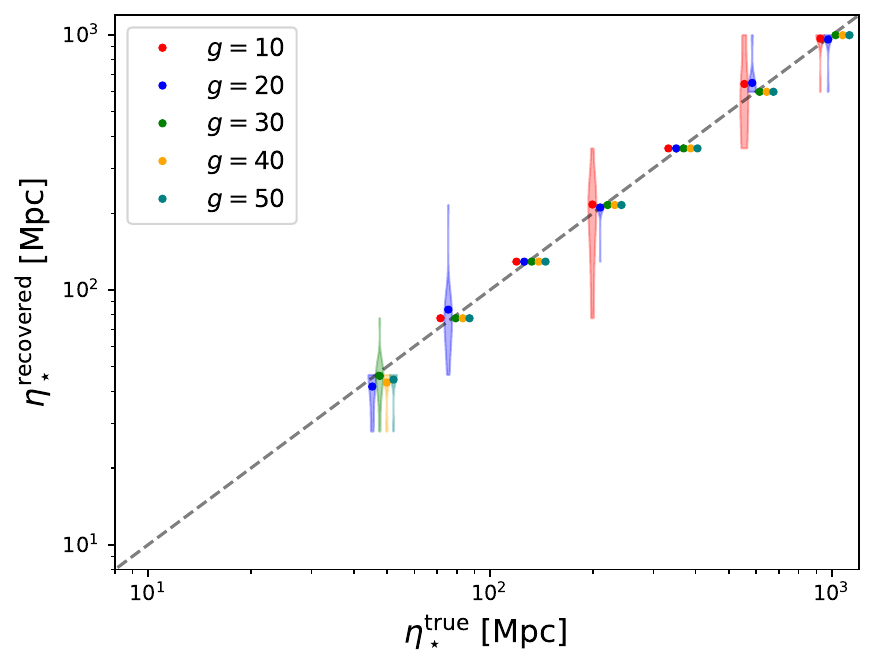}
  \hfill
  \includegraphics[width=0.32\textwidth]{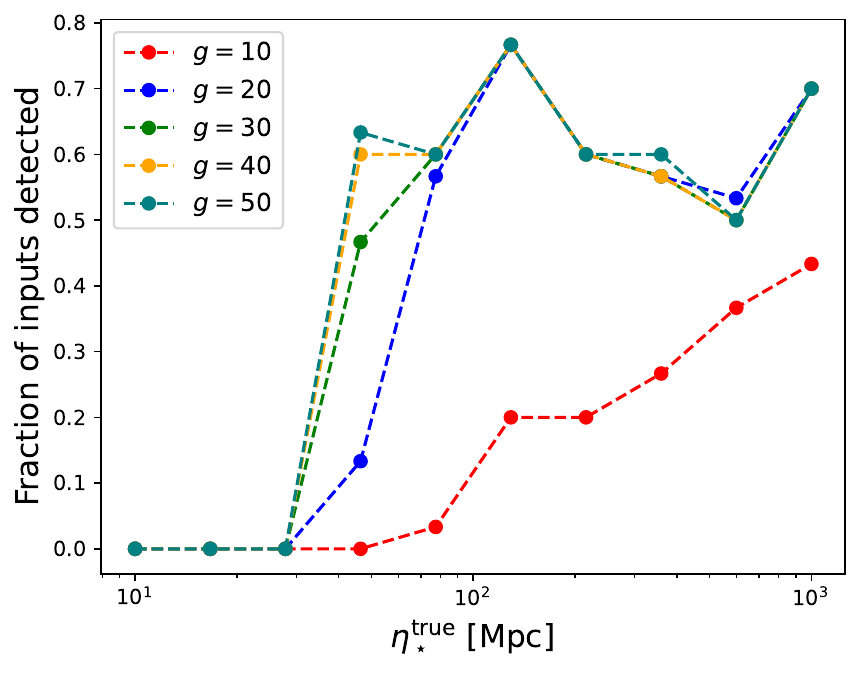}
  \hfill
  \includegraphics[width=0.32\textwidth]{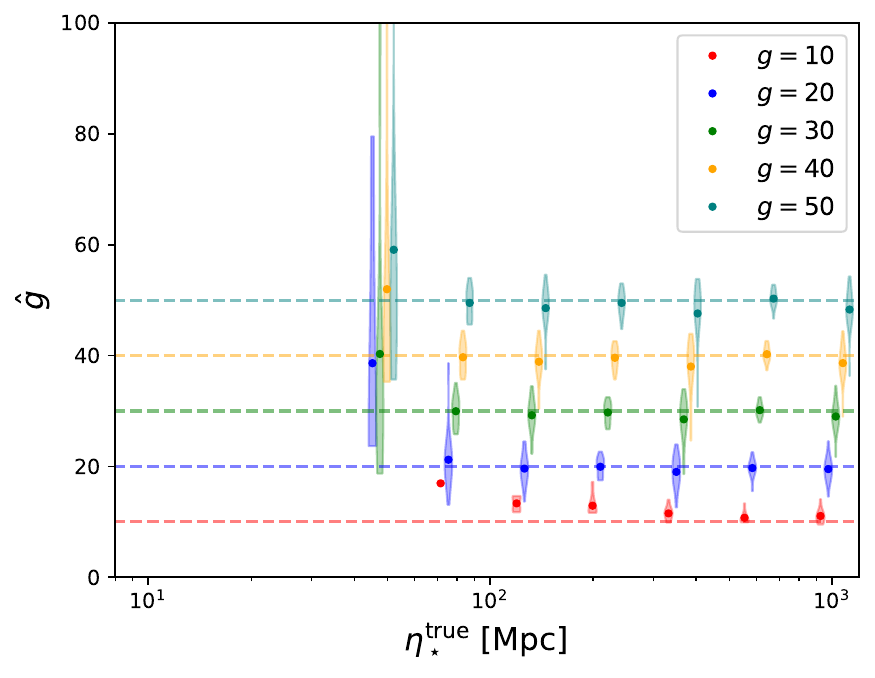}
  \caption{Summary of the analysis procedure detailed in Sec.~\ref{sec:meth} applied to the simulations detailed in Sec.~\ref{sec:sims}. \textbf{(Left)}: estimated values of $\eta_*$ from the injected hotspots. For sufficiently high $g$ and $\eta_*$, the recovered values reproduce the injected values more accurately among hotspots than found in the temperature search of~\cite{Philcox:2024jpd}. \textbf{(Central)}: fraction of injected sources recovered at SNR~$\geq 5$ as a function of the coupling parameter $g$ and hotspot size $\eta_*$. Note that due to masking (primarily of the Galaxy), our result  asymptotes to $\sim 60\%$; relative to the masked input catalog, the completeness is effectively 100\% for sufficiently large values of $g$ and $\eta_*$. Larger hotspots are easier to detect: we are limited on small scales by the \emph{Planck} beam and on large scales by the size of the tiles used in our search and by the Galactic mask. This selection function will be used to place constraints on the hotspot scenario in Sec.~\ref{sec:bounds}.  \textbf{(Right)}: recovered values of $g$ for the injected sources. Once again, for $g\geq20$, we recover $g$ accurately for sufficiently large hotspot sizes.} 
  \label{fig:widefigure_sims}
\end{figure*}

In Figure~\ref{fig:widefigure_sims}, we show the primary results of our simulation analysis. For sufficiently high $g$ and $\eta_*$, the completeness is very high, \textit{i.e.}, we detect the hotspots we inject. Note that because the mask leaves $\simeq60\%$ of the sky unmasked, we cannot obtain perfect completeness relative to the unmasked catalog. {We will return to this point in Sec.~\ref{sec:bounds} when we discuss the selection function}.   

In general, we find that the efficacy of our polarization search is consistent with that of the previous temperature search~\cite{Philcox:2024jpd}. 
Our search is sensitive to hotspots with $(g>20,\eta_*>25$~Mpc), for which the completeness ratio is $62.9\pm2.5\%$, matching that of~\cite{Philcox:2024jpd}~($64\pm7\%$). 
As for the temperature search, the completeness is poor for $g=10$ and does not exceed $\sim50\%$, even for large $\eta_*$. Note that this incompleteness means that if we hope to detect such hotspots, many of them must have been produced in the early universe.  As $\eta_*$ decreases (and thus the hotspots get smaller), consistent with~\cite{Philcox:2024jpd}, the completeness reduces significantly. We also present our pipeline's recovery of $g$ and $\eta_*$, finding that $\eta_*$ is well recovered for $\eta_* \gtrsim 50~\text{Mpc}$ and $g$ is successfully recovered for $\eta_* \gtrsim 100$~Mpc. This is consistent with the results of~\cite{Philcox:2024jpd}. 

Our simulation test results can be simply understood: for large $g$ the hotspots are brighter and easier to detect, while for large $\eta_*$, they are bigger and thus have no trouble being resolved by the \emph{Planck} beam.  Our conclusion of this simulation test is that our \emph{Planck} pipeline can robustly detect $E$-mode hotspots with {high completeness for} $g\geq20$ and $\eta_*\geq70~\text{Mpc}$. We also note that our $E$-mode search presents unbiased estimates on $(g,\eta_*)$ for $g\geq20$ and $\eta_*\geq70~\text{Mpc}$.

Many of the $g$ values discussed above would na\"ively break the perturbativity bounds of the calculation done in~\cite{Kim:2021ida}, which should require $g \ll \sqrt{4\pi}$. Generically, one would expect large radiative corrections when this bound is violated. In situations with large amounts of symmetry (e.g., supersymmetry~\cite{Kim:2021ida}) one can attain effective values of relatively high $g$. Such scenarios can also arise in extra-dimensional models, scenarios with tachyonic Higgs production, or due to quantum diffusion effects~\cite{ezquiaga2023massivegalaxyclusterslike,shakya2023tachyonichiggsinflationaryuniverse,kumar2025earlygalaxiesrareinflationary}. 

The low completeness for perturbatively small values of $g$ does not preclude our ability to constrain such parameter regimes; if the hotspots are numerous enough, we expect to detect a nonzero fraction of them as long as the completeness is non-zero. This discussion is continued and made more precise in Sec.~\ref{sec:bounds}.

\section{Searching for individual hotspots with \emph{Planck}}
\label{sec:Planck}

\noindent Next, we apply our pipeline to \emph{Planck} component-separated $E$-mode polarization maps using the common and point-source inpainting masks described in Sec.~\ref{sec:meth}. Our search uses 100 position-space templates with varying $\eta_*,
\eta_{{\mathrm{HS}}}$. For robustness, we apply the procedure to both \textsc{sevem} and \textsc{smica} component-separated maps~\cite{Planck:2020olo,Carron_2022,Philcox:2025wts}. Generating a candidate hotspot catalog using our matched-filter method takes $\sim 300~\text{CPU}$-hours. 

To avoid double-counting, we merge any two hotspots candidate within $\eta_*$ of one another. While this may seem dangerous, we adopt this approach for two reasons. First, this follows the convention used in tSZ cluster searches~\cite{Zubeldia:2022gva,Zubeldia_2023}. Second, bright candidates near the edges of our flat-space tiling are very likely to be detected in multiple tiles, and would otherwise be double-counted.  Additionally, because of the four free parameters of our template profile ($\eta_{{\mathrm{HS}}}$, $\eta_*$, and the angular coordinates of the hotspot center), when one finds a true source in a simulation, many candidates are detected locally in a small region around the injected hotspot. As such, our method avoids false over-detections. We further emphasize that this is an important reason why visual inspection plays a role in our final analysis. Because physically the hotspots are produced in pairs, upon the detection of a real hotspot, we should find a similar signal nearby (within distance $\eta_*$), which would likely have been merged in our analysis, but would remain visible to visual inspection.

\begin{figure*}[t]
  \centering
  \includegraphics[width=0.96\textwidth]{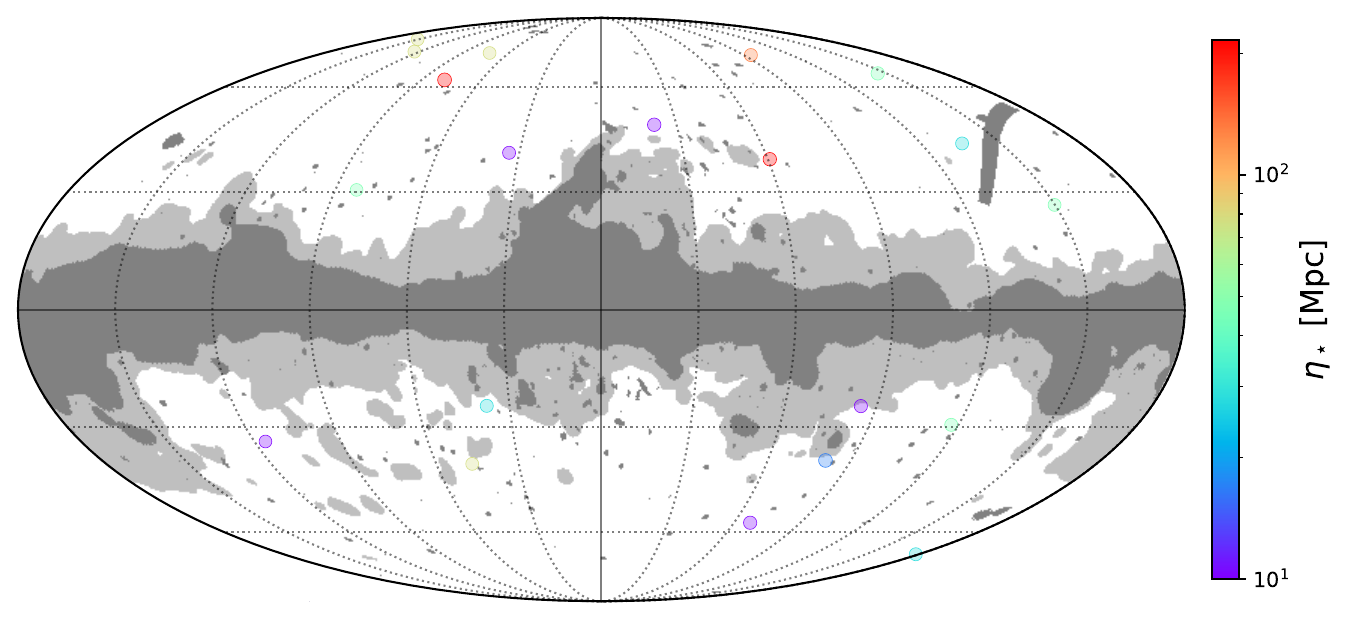}
  \caption{Visualization of the hotspot detection candidates from the \emph{Planck} \textsc{sevem} $E$-mode polarization maps. The dark and light shaded areas show the regions removed by the analysis mask and the post-processing mask, respectively. Each of the 20 circles represents a single detection candidate with SNR $\geq$ 5, whose size is proportional to the SNR and whose color indicates the inferred value of $\eta_*$. A histogram of the detection significances is shown in Figure~\ref{fig:snr}, with the parameters of all 20 of the SNR $\geq5$ detections given in Table~\ref{tab:pHScandidates}. We find no strong candidates. 
  }
  
  \label{fig:widefigure_scatter}
\end{figure*}

In Figure~\ref{fig:widefigure_scatter}, we show the sky distribution of hotspot candidates with SNR~$\geq5$. Notably, there is no obvious clustering of the spots in position. Also, due to the four free parameters, the SNR is actually not a true detection significance (\textit{i.e.}, SNR~$=5$ does not correspond to a 5$\sigma$ detection). As such, for a confident detection a more conservative threshold would be SNR~$\geq 6$, which was used in~\cite{Philcox:2024jpd}. 

\begin{figure}[t]
\includegraphics[width=0.48\textwidth]{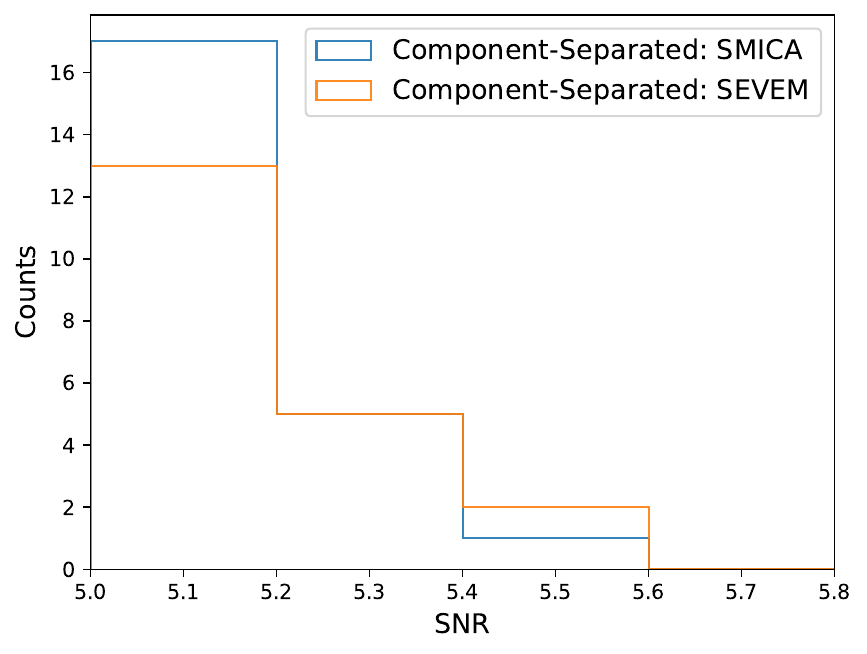}
\caption{\label{fig:snr} PDF of the SNR values of our $E$-mode hotspot candidates. The SNR is not a true significance due to the four free parameters in the template ($\eta_*$, $\eta_{{\mathrm{HS}}}$, and the hotspot coordinate center). We also emphasize the consistency between the \textsc{sevem} (orange) and \textsc{smica} (blue) results, which show similar hotspot numbers and SNR distributions.}
\end{figure}

In Figure~\ref{fig:snr} we show the distribution of SNRs for hotspot candidates. Applying the procedure to the \textsc{sevem} (\textsc{smica}) maps we find 20 (23) candidates after the application of our more conservative Galactic mask, with a maximum SNR of 5.4 (5.5). The distributions of SNR values are consistent between the two maps.

\begin{table}[!t]
\centering
\begin{tabular}{cccccc}
\hline
   SNR &      Longitude [$^\circ$] &      Latitude [$^\circ$] &        $\hat{g}$ &   $\eta_*$ [Mpc] &   $\eta_{{\mathrm{HS}}}$ [Mpc] \\
\hline
 5.0  & 150.7  &  77.5 &   $20^*$ &    77.4 &  217.1 \\
 5.0 &  72.5 &  72.0 &   17 &    77.4 &  245.2 \\
 5.1 & 123.5  &  72.5 &   16 &    77.4 &  259.3 \\
 5.3 & 214.6  &  64.6 &   $29^*$ &    46.4 &  259.3 \\
 5.1 & 265.4  &  71.2 &   12 &   129.2  &  292.2 \\
 5.4 &  78.5 &  62.4  &   16  &   215.4  &  378.4 \\
 5.1 & 224.6  &  43.1 &   $83^*$ &    27.8 &  293.1 \\
 5.3 & 338.9  &  48.5 & $1556^*$   &    10.0      &  272.2 \\
 5.3  & 299.3  &  38.8 &   14 &   215.4  &  339.2 \\
 5.1 &  33.7 &  40.5 & $1246^* $  &    10.0      &  283.2 \\
 5.0 &  82.8   &  30.5 &   25  &    46.4 &  242.4 \\
 5.0 & 210.2  &  26.6 &   $31^*$ &    46.4 &  259.3 \\
 5.1 &  37.4  & -24.4 &   $77^*$ &    27.8 &  272.8 \\
 5.2 & 275.1  & -24.4 &  $801^*$  &   10.0   &  279.5\\
 5.1 & 242.5  & -29.3 &   $32^*$&    46.4 &  318.4 \\
 5.0 & 116.2  & -33.7  & $1160^*$   &    10.0 &  272.2 \\
 5.4 & 279.2  & -38.9 &  187  &    16.7  &  266.8 \\
 5.1 &  46.9 & -39.8 &   21 &    77.4 &  329.7 \\
 5.2 & 292.7  & -57.1 &  $885^*$   &    10.0 &  272.2 \\
 5.1 & 181.8  & -67.5 &   $93^* $&    27.8&  293.1 \\
\hline
\end{tabular}
\caption{Inferred parameters for hotspot candidates with SNR~$\geq 5$ from the \emph{Planck} \textsc{sevem} component-separated $E$-mode maps. Any of the candidates marked with $^*$ are within the parameter space covered by the temperature search~\cite{Philcox:2024jpd}, but were not detected in the former work. For the remaining eight candidates, all of them are either at low enough $\eta_*$ or $g$ that our matched-filter pipeline cannot reliably recover hotspots because the \emph{Planck} beam is too large, as shown in Figure~\ref{fig:widefigure_sims}. By contrast, in the \textsc{sevem} temperature component-separated maps, \cite{Philcox:2024jpd} found 48 candidates with SNR~$\geq5$, and SNR values up to 8. This may be due to stronger foregrounds in temperature than in polarization, in particular given that the high-SNR candidates were matched to point sources, and masking effects.} 
\label{tab:pHScandidates}
\end{table}

We find fewer candidates than found in the temperature analysis of~\cite{Philcox:2024jpd}: 20 (23) $E$-mode versus 48 (35) $T$ candidates for \textsc{sevem} (\textsc{smica}).  Notably, there are no candidates above an SNR of 6. We expect that many of the SNR $\geq 5$ candidates are random fluctuations (we present a visual inspection below in Appendix~\ref{app:vis}, which appears consistent with noise).  We also present in Table~\ref{tab:pHScandidates} the recovered parameters for each of the \textsc{sevem} $E$-mode candidates; as shown in Appendix~\ref{app:smica}, we find similar inferred parameters when analyzing the \textsc{smica} maps. Most of the hotspot candidates appear near the edge of masked regions (see Appendix~\ref{app:vis}), so we unsurprisingly find similar candidate locations in the \textsc{sevem} and \textsc{smica} maps, as well as similar recovered candidates. Twelve of these hotspots appear in regions of parameter space excluded by the temperature search, and as such we would expect them to have been seen in~\cite{Philcox:2024jpd}. As such, they are unlikely to be physical, though they are still visually inspected in Appendix~\ref{app:vis} as an additional cross-check. The remaining eight would likely not have been detected in the temperature search due to large scatter at the relevant $\eta_*$ seen in Sec.~\ref{sec:sims}. In summary, we find no hotspots with SNR~$>6$, and none with SNR~$>5$ with a distinctive profile that is obviously consistent with a primordial hotspot.

\begin{figure*}[t!]
  \centering
  \includegraphics[width=0.48\textwidth]{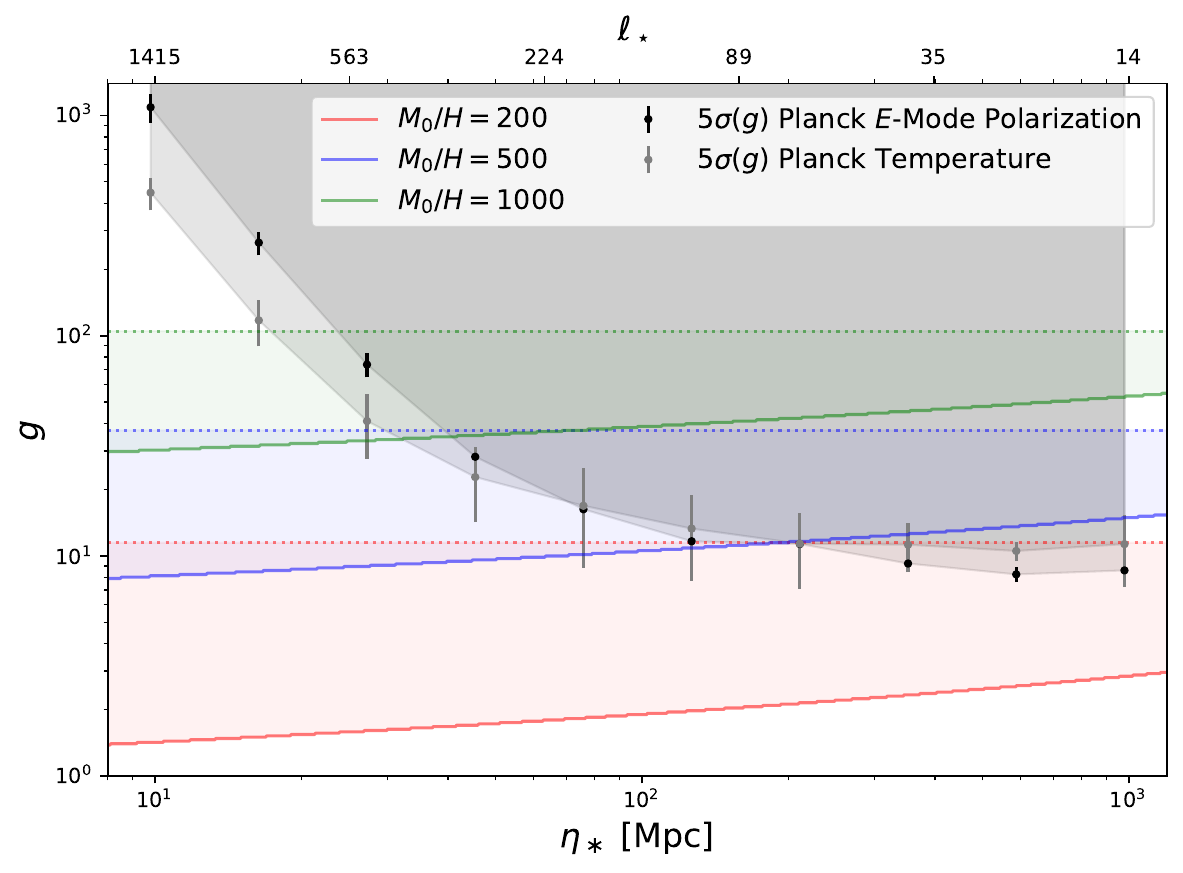}
  \hfill
  \includegraphics[width=0.48\textwidth]{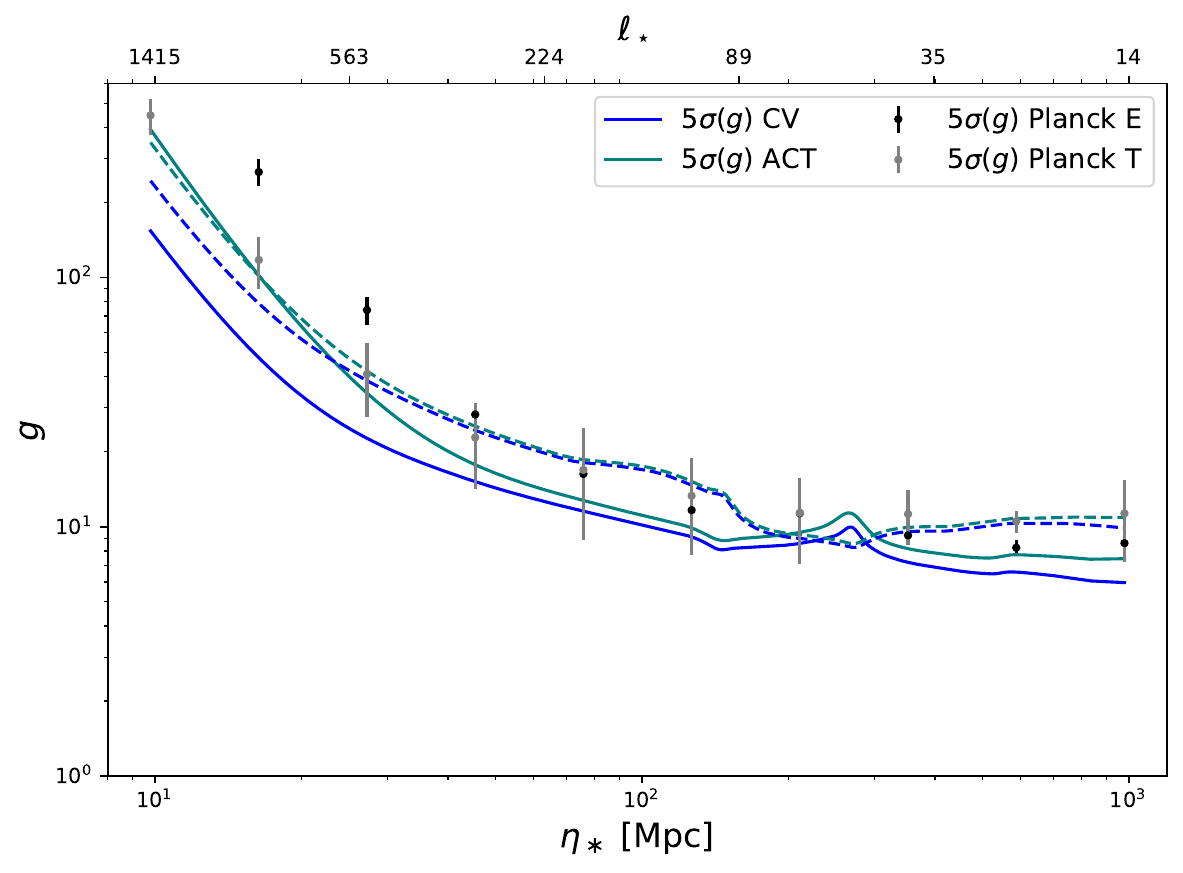}
  \caption{\textbf{(Left)}: Sensitivity to hotspots as inferred from the analysis of the \emph{Planck} \textsc{sevem} component-separated $E$-mode polarization data. The data points show $5\sigma$ error bars attained from the simulation in Sec.~\ref{sec:sims}, \textit{i.e.}, the shaded region above the data points is that for which we would reliably expect to detect a single hotspot at $5\sigma$. The colored regions indicate reasonable regions of parameter space corresponding to different values of the particle mass at particle production $M_0$. The solid line lower bound is set by requiring that at least one hotspot be produced (see Equation~\eqref{eqn:NHS}), and the upper limit is set by backreaction constraints from \cite{Kim:2021ida}. We also plot on the top axis the characteristic scale $\ell_*$, estimated by $\ell_*\simeq \eta_0/\eta_*$. Note that this is an approximate scale for the profile in harmonic space, \emph{not} the scale at which the signal-to-noise is maximized.  \textbf{(Right)}: We show forecast results for both ACT and an optimal cosmic-variance-limited experiment, where $E$ ($T$)-modes are represented by solid (dashed) lines. These forecasts are computed analytically using Equation~\eqref{eq:MMF}. Notice that for $T$, ACT comes very close to the CV limit for $\eta_*\geq30~\text{Mpc}$, and for $E$, ACT improves on the \emph{Planck} $E$-mode {sensitivity} by between a factor of $1.2-2.75$.}
  \label{fig:exclusions}
\end{figure*}

\section{Forecasting Sensitivity for Localized Hotspot Searches in Temperature and Polarization}
\label{sec:forecasts}

\noindent In the following we explore our matched filter pipeline's sensitivity to individual hotspots (the parameter regime in which we expect to robustly detect a single hotspot), and outline our formalism for deriving these sensitivities. The main results are contained in Figure~\ref{fig:exclusions}, where the left panel contains the simulation-derived \emph{Planck} $T$ and $E$-mode sensitivities, and the right panel compares them to forecast results from ACT and a CV-limited experiment, which is noise-free up to $\ell_{\rm{max}}=3500$. The left panel of Figure~\ref{fig:exclusions} makes clear that our polarization hotspot search provides independent information from that obtained using temperature data in~\cite{Philcox:2024jpd}. In particular, we see improvement in the sensitivity to hotspots, which is most pronounced for large $\eta_*$: we find stronger sensitivity to particle production from \emph{Planck} polarization data than from temperature data for $\eta_* \gtrsim 100$~Mpc.

Here, we detail our procedure for analytically computing a sensitivity forecast for both $T$ and $E$-mode hotspot profiles.  For simplicity, we fix $\eta_{\rm HS} = \eta_{\rm rec}$ in these forecasts.  Our approach follows from Equation~\eqref{eq:MMF}, though we work in the full-sky limit  for computational efficiency (with $t(\mathbf{l})\to t_{\ell m}$).  This can be done due to the isotropy of our profile, which allows us to reduce our problem to calculating $t_\mathrm{\ell0} \equiv t_\ell$.  From our definition of the position-space profiles in Equation~\eqref{eq:profile}, with $X \in \{T,E\}$, we can derive the harmonic space templates using the $m=0$ spherical harmonics $Y_{\ell0}=\sqrt{(2\ell+1)/4\pi}\,P_\ell$. Defining \begin{align}
   \mathcal{I}^X_\ell\equiv\int_0^{\infty}\frac{\dd k}{k} f(k\eta_*)\mathcal{T}_\ell^X(k) j_\ell(k\chi_{{\mathrm{HS}}}) \,,
\end{align}
we have: 
\begin{align}
    t^X_\ell& =&&\int_{S^2}\dd^2 \mathbf{\hat{n}}~ Y_{\ell0}^*~\delta X(\theta,\eta_*,\eta_{{\mathrm{HS}}})\nonumber\\
    &=&&\sum_{\ell'=2}^{\infty}\sqrt{\frac{(2\ell+1)^3}{4\pi^3}}\mathcal{I}^X_{\ell'}\int_{-1}^1\dd \cos\theta P_\ell(\cos{\theta})P_{\ell'}(\cos{\theta})\nonumber\\
    &=&&
    \sqrt{\frac{2\mathrm{\ell}+1}{\pi^3}}\int_0^{\infty}\frac{\dd k}{k} f(k\eta_*)\mathcal{T}_\ell^X(k) j_\ell(k\chi_{{\mathrm{HS}}}).
    \label{eq:harmonicprofs}
\end{align}
In the third equality we have used the orthogonality of Legendre polynomials, $\int_{-1}^1 \dd xP_\ell(x) P_{\ell'}(x)=2/(2\ell+1)\delta_{\ell\ell'}^{\rm K}$.
We can then compute the variance of the matched-filter estimator: 
\begin{align}
\label{eq.var}
(\sigma_g^X)^{-2}=\sum_{\ell}\frac{[t_\mathrm{\ell}^X(\eta_*,\chi_{{\mathrm{HS}}})]^2}{C_{\ell,\text{Exp}}^{XX}} \,,
\end{align}
where $C_{\ell,\text{Exp}}^{XX}=C_{\ell,\Lambda\text{CDM}}^{XX} + \mathcal{N}^{XX}_{\ell,\text{Exp}}$ denotes the beam-deconvolved power spectrum for a given experiment, including the noise power spectrum $\mathcal{N}^{XX}_{\ell,\text{Exp}}$. As in our \emph{Planck} data analysis, we define the exclusion curve at SNR = $5$ assuming a null hotspot signal, \textit{i.e.}, we consider the bound $g\geq 5\sigma_g$. 

It is computationally straightforward to extend this approach to a joint analysis of $T$ and $E$, by analogy to the multi-frequency matched-filter formalism~\cite{Haehnelt:1995dg,Melin:2006qq,Zubeldia_2023,Zubeldia:2022gva}.  In the joint analysis, we must account for the non-zero covariance of the $T$ and $E$ fields arising from $C_{\ell}^{TE}$.  The variance of the resulting joint estimator is given by 
\begin{align}
\label{eq.TEvar}
    \left(\sigma_g^{T\times E}\right)^{-2} = \sum_{X,X'\in\{T,E\}}\sum_{\ell}t_{\ell}^{X} C_{\ell,\text{Exp}}^{-1,XX'} t_{\ell}^{X'} \,.
\end{align}
Here, $C_{\ell,\text{Exp}}^{-1,XX'}$ is the matrix inverse of the full $(T,E)$ covariance matrix, \textit{i.e.}, the matrix containing diagonal blocks of $C_\ell^{TT}$ and $C_\ell^{EE}$ and an off-diagonal block of $C_\ell^{TE}$. In practice, while this joint $T\times E$ analysis would yield minor improvements for the \emph{Planck} sensitivities ($10-20\%$), they are not nearly as striking as the improvement associated with using higher-resolution CMB data (for which one could well do a joint analysis). 

We now apply this procedure to an optimal cosmic-variance-limited experiment (denoted CV, with $\mathcal{N}_\ell=0$), with results shown in Figure~\ref{fig:exclusions}.  We find dramatic differences between the sensitivities from temperature and polarization, where for $\eta_*\leq 150~\text{Mpc}$ or $\eta_*\geq 500~\text{Mpc}$ the sensitivities from $E$ are $\sim1.6\times$ better than those from $T$, and on intermediate scales the improvement factor varies between 0.8 (in the small region of parameter space where $T$ is superior) to 1.4. Strikingly, polarization provides almost uniformly stronger sensitivity than temperature across the full parameter space. This is emphasized by the right panel of  Figure~\ref{fig:exclusions}, where we see that for $\eta_*\geq 300$~\text{Mpc}, \emph{Planck} E-mode hotspot sensitivities are stronger than those inferred from a CV-limited temperature experiment.

We should also note that the CV limit is very sensitive to the value of $\ell_{\rm max}$ used in the sum in Equations~\eqref{eq.var} and~\eqref{eq.TEvar}.  In Figure~\ref{fig:widefigure_FISHforcs} (black curves), we present results for $\ell_{\rm max}=3500$, which provides a reasonable comparison with current and proposed future experiments~\cite{Atkins:2024jlo,ACT:2025fju,Bleem_2022,SimonsObservatory:2025wwn,Abazajian:2019eic}.  We also demonstrate in Figure~\ref{fig:widefigure_FISHforcs} that in the optimal case a joint $T \times E$ search  only provides marginal improvement over a search using polarization alone ($\sim 5-25\%$, where the maximum is reached in the small regime where temperature constraints overtake those from polarization). This occurs since the temperature and polarization are only weakly correlated, and the constraints due to polarization are almost uniformly better than those from temperature. The fact that polarization dominates the constraints is quite interesting and is a striking consequence of the differing noise (including noise from the standard-model CMB fluctuations) and transfer functions in $T$ and $E$.

Next, we directly forecast the sensitivity of the Atacama Cosmology Telescope (ACT)~\cite{Atkins:2024jlo,ACT:2025xdm,ACT:2025fju,ACT:2025tim,Beringue:2025bur} to individual inflationary hotspots.  For this forecast, we use the beam-deconvolved noise power spectra for both temperature and $E$-mode polarization from ACT Data Release 6~\cite{ACT:2025fju}. Our forecast suggests that the bounds of this work can be significantly enhanced with the ACT data  on small scales (see Figure~\ref{fig:exclusions}).  For temperature, ACT should nearly reach the CV limit (at $\eta_*\gtrsim30~\text{Mpc}$). We also find that polarization provides significantly stronger bounds on intermediate scales  ($25~\text{Mpc} \lesssim \eta_* \lesssim 300~\text{Mpc}$) than temperature, by a factor of $\sim 1.25-1.5$. For large $\eta_*$, we find that \emph{Planck} constraints are comparable to those forecast for ACT; this is not surprising, since $\eta_*=1000~\text{Mpc}$ corresponds to a characteristic $\ell_*=\eta_0/{\eta_*}\simeq14$, a scale sufficiently large that atmospheric noise renders the ACT data much less powerful than \emph{Planck}~\cite{ACT:2025xdm,ACT:2025fju}. In general, our forecasts (Figure \ref{fig:exclusions}) indicate that ACT will provide improved bounds on extremely massive particle production during inflation.  It should also be noted that very small hotspots are generically difficult to constrain because of finite resolution effects from the beam.

We should emphasize that in this section we have considered the sensitivity of our pipeline to a \emph{single} hotspot in different regions of parameter space, comparing $T$ and $E$-mode data. We can then map these sensitivities onto constraints on the underlying parameters of the physical theory, by taking into account the number of predicted hotspots at each point in parameter space. In the following section we will build a Poissonian likelihood to obtain such constraints.

\section{From Sensitivity to Parameter Constraints}
\label{sec:bounds}

\noindent To derive constraints on the underlying physical model of hotspot production, we build a Poissonian (Cash-statistic~\cite{1979ApJ...228..939C}) likelihood analogous to that used in cluster abundance cosmology~\cite[e.g.,][]{2001ApJ...560L.111H,Hasselfield:2013wf,Planck:2015lwi,SPT:2024qbr}: 
\begin{align}
   -2 \ln \mathcal{L}(\vec{p}) = -2 \left( N_{\text{obs}} \ln(N_{\text{pred}}(\vec{p})) - N_{\text{pred}}(\vec{p}) - \ln(N_{\text{obs}}!) \right) \,,
\end{align}
where $N_{\rm obs}$ ($N_{\rm pred}$) is the observed (predicted) number of hotspots in a given CMB survey, and the latter is a function of the model parameters $\vec{p} \equiv \{g, M_0, \eta_*\}$.  Note that $\eta_{\rm HS}$ only enters the theory model via the selection function, and it will thus be integrated over (see below).  In the case of a null detection (as in our analysis), we have $N_{\rm obs} = 0$ and thus
\begin{align}\label{eq:likelihood_simple}
    \ln \mathcal{L}(\vec{p}) = -N_{\text{pred}}(\vec{p}) \,.
\end{align}

\noindent The predicted number of hotspots is given by 
\begin{align}\label{eq:Npred}
    N_{\text{pred}}(g,M_0,\eta_*) = \left(\frac{1}{H_I\eta_*}\right)^3 \int \dd^3 x \, n(g,M_0) \,\mathcal{S}(g,\eta_*,\eta_{\text{HS}}) \,,
\end{align}
where the predicted number density is (cf.~Eq.~\eqref{eqn:NHS})
\begin{align}\label{eqn:number}
    n(g,M_0) =\left( \frac{g\dot{\phi}_0}{4\pi^2} \right)^{\frac{3}{2}}\exp{-\frac{\pi( M_0^2-2H_I^2)}{g\dot{\phi}}},
\end{align}
and $\mathcal{S}(g,\eta_*,\eta_{\text{HS}})$ is the selection function, \textit{i.e.}\ the probability of detecting an individual hotspot with parameters $(g,\eta_*,\eta_{\text{HS}})$.  Note that the selection function depends on the specifications of a given experiment, including its sky coverage.  For example, the maximum possible detection probability, even for a very bright hotspot, is $f_{\rm sky}$, the observed sky fraction.  Note also that the selection function depends on $\eta_{\rm HS}$, but the predicted number density of hotspots does not.  In Eq.~\eqref{eq:Npred}, the integral over comoving volume integrates over all possible hotspot locations.

A simulation-based estimate of the selection function for our \emph{Planck} E-mode analysis is shown in the central panel of Fig.~\ref{fig:widefigure_sims}.  However, in our case the selection function can be modeled analytically using the uncertainty associated with our matched filter estimator.  Treating the output of the matched filter as a Gaussian random variable, it is straightforward to show that the selection function (for a $5\sigma$ detection) takes the form\footnote{The only major assumption we make here is that we may consider the selection function averaged over the full sky, rather than on a tile-by-tile basis.  Given the relatively homogeneous noise in \emph{Planck}, this is a reasonable approximation.}~\cite{Planck:2015vgm}
\begin{align}
    \mathcal{S}(g,\eta_*,\eta_{HS})=\frac{f_{\text{sky}}}{2}\text{erfc}\left(\frac{1}{\sqrt{2}}\left(5-\frac{g}{\sigma_g}\right)\right) \,,
    \label{eq:selection_function}
\end{align}
where $\text{erfc}$ is the complementary Gaussian error function, $\text{erfc}(x)\equiv1-\frac{2}{\sqrt{\pi}}\int_0^xe^{-x'^2/2}\dd x'=\int_x^{\infty} e^{-x'^2/2}\dd x'$, and $\sigma_g$ is given by Eq.~\eqref{eq.var}. Note that all of the selection function's dependence on $\eta_*$ and $\eta_{\text{HS}}$ is contained in $\sigma_g$. We provide a brief derivation of this result in Appendix~\ref{Appendix:appendix D}. Fig.~\ref{fig:selectionplot} demonstrates that the analytically computed selection function agrees within $1\sigma$ with the simulation-derived selection function, where the latter is computed as discussed in Section~\ref{sec:sims}, by injecting 30 hotspots into simulated sky maps for each $\eta_*$ value in our pipeline.

We perform a likelihood analysis using the formalism described above. Given the strong degeneracy between $g,\eta_*$, and $M_0$, we first fix characteristic values of $M_0$.  For each value of $M_0$, we consider ten logarithmically spaced values of $\eta_*$, each of which is used to perform a one-parameter likelihood analysis for $g$. We then present constraints in the $(g,\eta_*)$ plane for each $M_0$ value. Given the simple nature of our one-parameter likelihood, we directly compute the $95\%$ confidence upper limit on $g$ by computing $\Delta \chi^2=-2\ln{\mathcal{L}}$ and noting that the maximum-likelihood point is at $g=0$ (simply from the form of our likelihood in Eq.~\eqref{eq:likelihood_simple} with no observed hotspots). The one-sided confidence interval is then determined by finding value of $g$ such that $\Delta \chi^2(g)=3.84$.

\begin{figure*}[t!]
  \centering
  \includegraphics[width=0.48\textwidth]{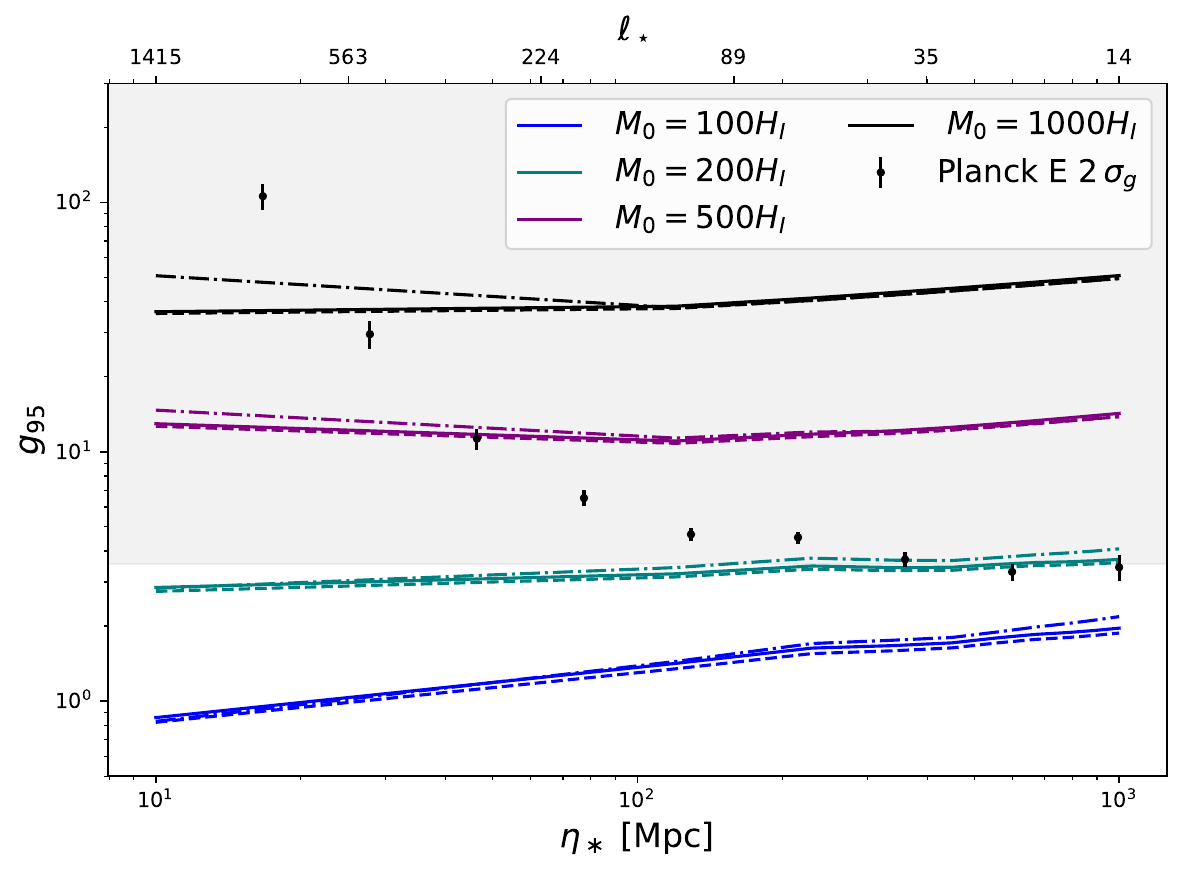}
  \hfill
  \includegraphics[width=0.48\textwidth]{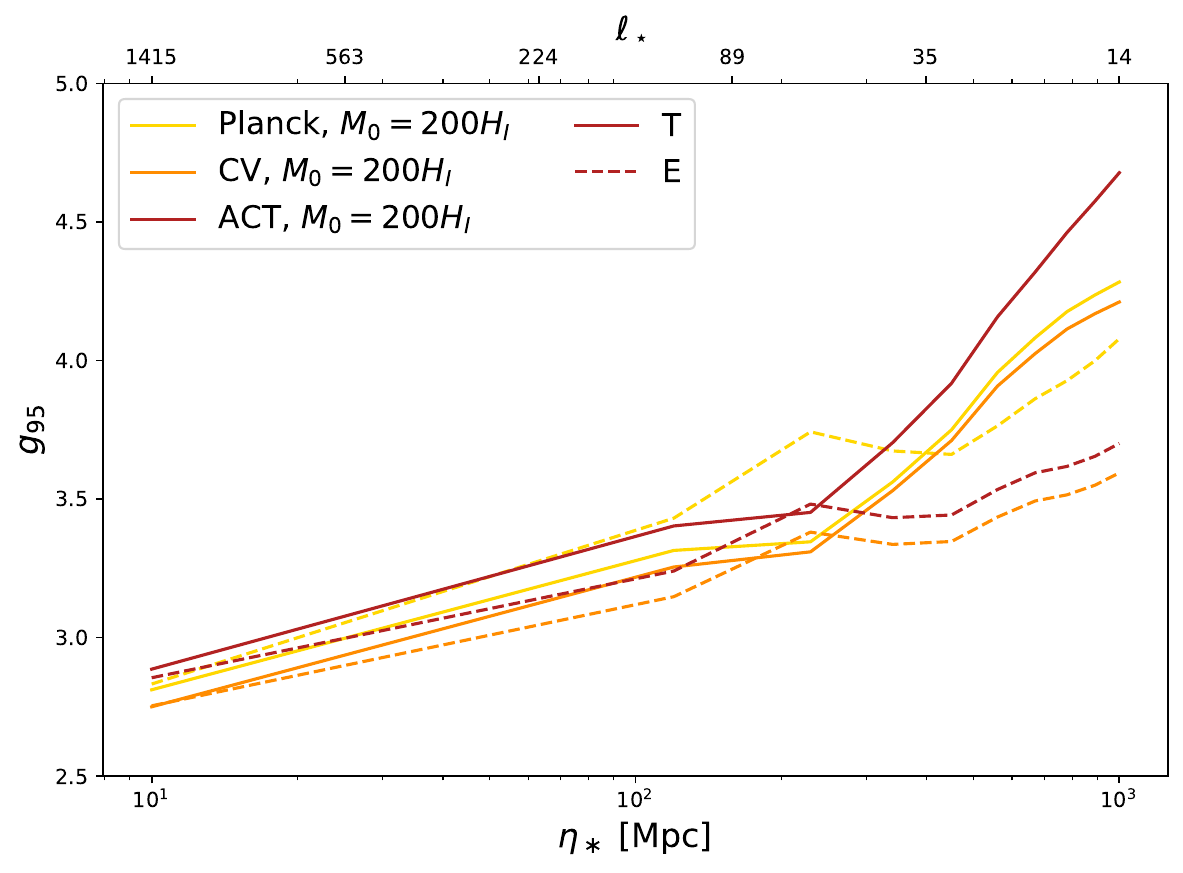}
  \caption{\textbf{Left}: Exclusion plot inferred from the analysis of the \emph{Planck} \textsc{sevem} component-separated $E$-mode polarization data. The data points show $95\%$ confidence upper limits on the coupling $g$ derived from our Poissonian likelihood in Sec.~\ref{sec:bounds}, which rely on the instrumental sensitivities computed in Sec.~\ref{sec:forecasts}. Dashed lines represent a cosmic-variance-limited experiment up to $\ell_{\rm max}=3500$ with $f_{\text{sky}}=0.6$, solid lines denote bounds that would be inferred from a non-detection with ACT, and the dot-dashed lines show our \emph{Planck} bounds. Note that for lower-mass particles ($M_0 \lesssim 200H_I$, we constrain values of the coupling within the perturbative regime (which is indicated by the shaded gray region). \textbf{Right}: Exclusion bounds inferred from \emph{Planck}, ACT, and a CV-limited experiment in both $T$ (solid) and $E$ (dashed), fixing $M_0=200H_I$. We note a couple of important features. First, $E$-mode data dominate the constraints in both ACT and the CV-limited experiment. Second, \emph{Planck} provides tighter constraints on small angular scales (for this choice of $M_0$) due to ACT's smaller sky coverage (despite the improved sensitivity per mode). At even smaller $\eta_*$ values than shown here, ACT would eventually take over as the resolution limit of \emph{Planck} is reached.}
  \label{fig:trueexclusions}
\end{figure*}

In Fig.~\ref{fig:trueexclusions}, we present constraints obtained from the \emph{Planck} search.  We also show forecast constraints for ACT (assuming a null detection), where the only additional input beyond the sensitivities derived in Sec.~\ref{sec:forecasts} is the value of $f_{\rm{sky}}$, which we take to be $f_{\rm{sky}}=0.35$. Finally, we consider a CV-limited experiment, with the same sky coverage as our \emph{Planck} search ($f_{\rm{sky}}=0.6$). As expected from  the form of our Poissonian likelihood, our constraints worsen with $M_0$ due to the exponential suppression of $N_{\text{pred}}$ with $M_0^2$.  Conversely, in the lower mass range corresponding to large $N_{\rm{pred}}$, we find strong bounds on $g$.  In Fig.~\ref{fig:trueexclusions}, we see that for theories with $M_0=100H_I$, the \emph{Planck} search bounds $g\leq 2$ at 95\%~C.L., for any $\eta_* \leq 1000$~Mpc. Thus, our \emph{Planck} constraints from the full Poissonian likelihood are at least a factor of two stronger than the suboptimal bounds derived only from the single-hotspot sensitivities (\textit{e.g.,} as computed in Ref.~\cite{Philcox:2024jpd}), which we show for the $E$-mode analysis in black points in the figure.  In the low $\eta_*$ regime, where many hotspots are produced but individually they are difficult to detect, we see an improvement of nearly three orders of magnitude over the single-hotspot sensitivity.

More generally, we obtain much stronger constraints with the full likelihood than the sensitivity-based approach: our constraints reach the perturbative regime for $M_0 \lesssim 500H_I$, while the single-hotspot sensitivity bounds na\"ively exceed the perturbativity bound for low $\eta_*$.\footnote{Appendix~D of \cite{kumar2025earlygalaxiesrareinflationary} shows that the observed coupling can be modified if our theory is inherited from a 5D model, allowing larger (formally non-perturbative) values of $g$ to be obtained.}  Nevertheless, for sufficiently large $\eta_*$, where the expected number of produced hotspots becomes less than one on the full sky, we see the sensitivity-based bounds surpass those from the full-likelihood.  However, we should emphasize that the weaker likelihood-based constraints in this regime are both expected and correct. If we do not expect any hotspots to be produced in some region of parameter space, we cannot constrain the physical model in that regime, no matter how reliably we would expect to detect an individual hotspot. 
Overall, our \emph{Planck} search bounds particles with $M_0/H_I=(100,200, 500,1000)$ to have $g \leq (2,4,15,55)$ at 95\%~C.L.~across all scales, and in the low $\eta_*$ regime, particles with $M_0=100H_I$ are constrained to have couplings $g\leq1$.

It is also worth discussing how the \emph{Planck} bounds derived in this work compare to those expected from current and future CMB experiments. We consider the full likelihood approach applied to the forecast sensitivities of the experiments considered in Sec.~\ref{sec:forecasts}. On these scales, we expect ACT $E$-mode data to improve on our bounds by around $10\%$.\footnote{Note that if one pushed to smaller $\eta_*$ values and hence higher $\ell_{\rm{max}}$, ACT would continue to improve more substantially upon \emph{Planck}. 
There is a complicated interplay of scales; for low-mass particles (which have large abundances) \emph{Planck}'s improvement in sky fraction over ACT wins out, whereas for higher values of $\eta_*$ (where we expect fewer particles to be produced) ACT's stronger $E$-mode sensitivity dominates. Naturally, the CV-limited experiment provides even stronger bounds ($\sim 15\%$). In sum, on the scales considered in this work, one can make notable improvements upon the \emph{Planck} results, and we expect that using higher-resolution CMB data we would be able to probe even smaller $\eta_*$ values.}
Additionally, the right panel of Fig.~\ref{fig:trueexclusions} provides a comparison of the potential to derive constraints from $T$ and $E$-mode data jointly. As expected from Fig.~\ref{fig:exclusions}, we find that $E$-modes dominate the constraints across nearly all scales when using ACT or CV noise properties (with $\ell_{\rm max}=3500$), while the $T$ constraints remain dominant in \emph{Planck}.

\section{Power Spectrum Searches}
\label{sec:powerspec}

\noindent Inflationary massive particles also induce changes to the CMB two-point function, which provide an alternative observational probe. Here we provide a derivation\footnote{We make the simplifying assumption that single hotspots are produced in an uncorrelated manner. Of course, from momentum conservation, hotspots are produced pairwise, and as such are correlated on distance scales $\alt \eta_*$. This effect is however subleading and could be explored in further work.} of the induced correction to the power spectrum from massive particle production (see also Appendix~C of Ref.~\cite{kumar2025earlygalaxiesrareinflationary}). Here we consider an approximate treatment, and leave a more careful analysis to future work.  We then compare the bounds inferred from a power spectrum-based search to those from a matched-filter search. Throughout, we consider an optimistic power spectrum search, considering a CV-limited experiment without experimental noise up to $\ell_{\rm{max}}=3500$, without marginalizing over $\Lambda$CDM parameters.

\begin{figure*}[t!]
  \centering
\begin{minipage}
{0.48\textwidth}
    \centering
\includegraphics[width=\linewidth]{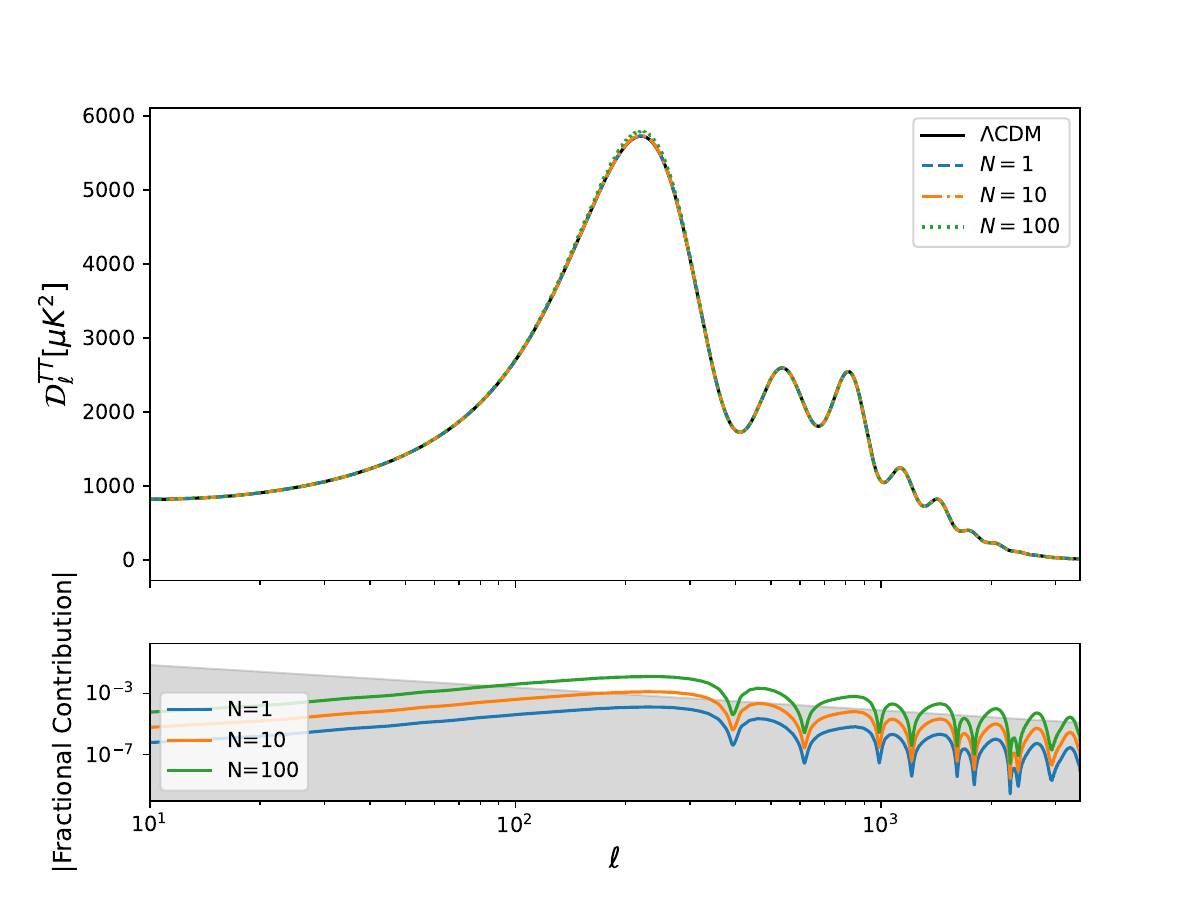}

{\textbf{A)~}\small $TT$ power spectrum.}
\end{minipage}  
  \hfill
  \begin{minipage}
  {0.48\textwidth}
    \centering
      \includegraphics[width=\linewidth]{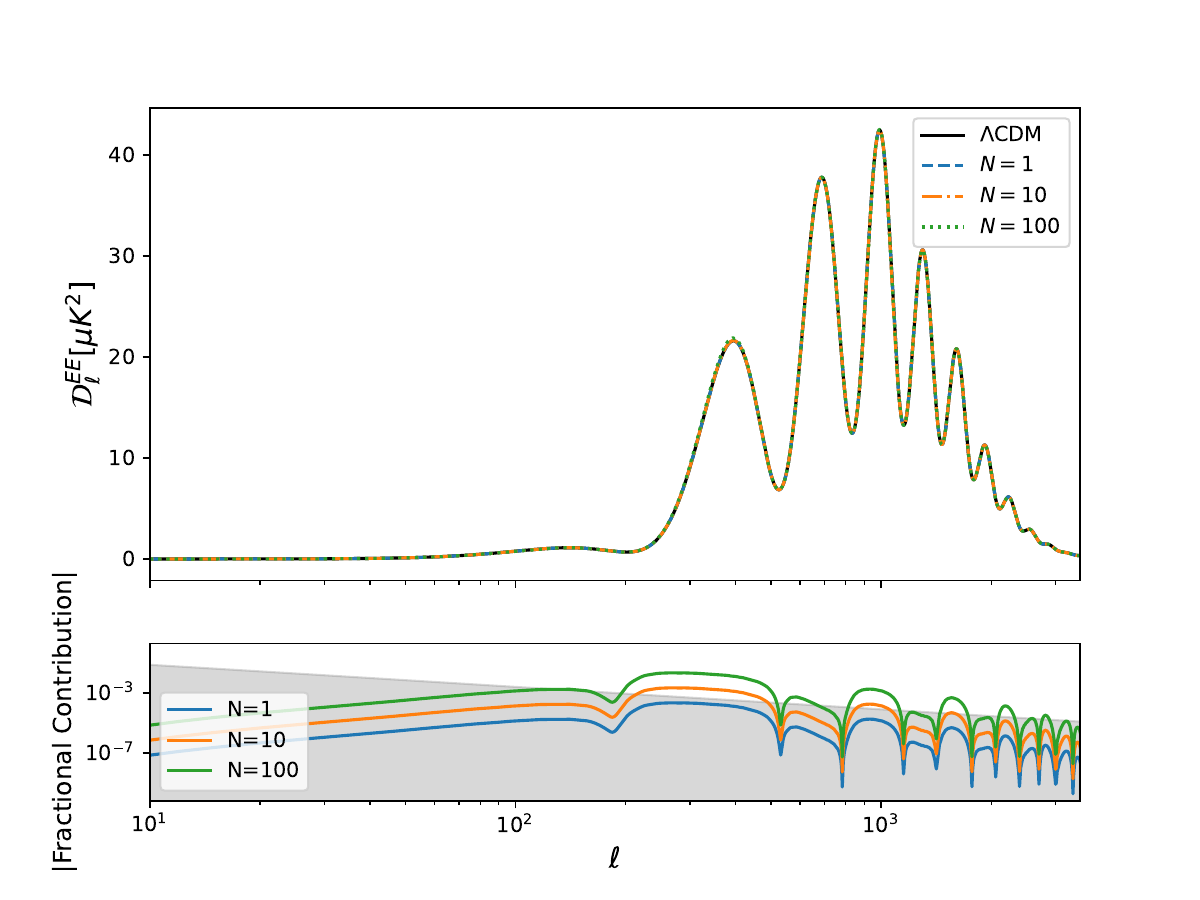}
      {\textbf{B)}~\small $EE$ power spectrum.}
  \end{minipage}
  \vspace{0.4cm}
  \begin{minipage}
    {0.48\textwidth}
    \centering\includegraphics[width=\linewidth]{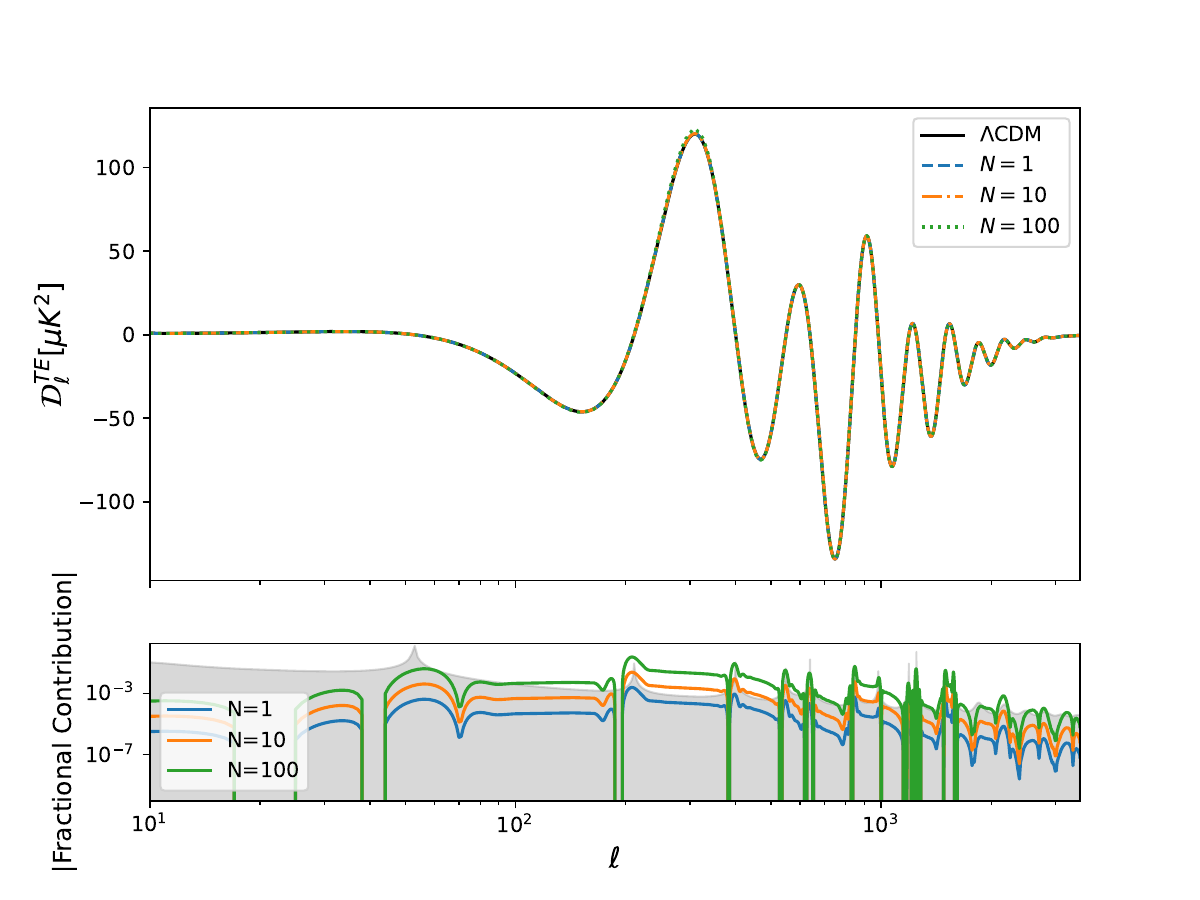}
    {\textbf{C)}~\small $TE$ power spectrum.}
  \end{minipage}
  \caption{Power spectra for $TT$ (top left), $EE$ (top right), and $TE$ (bottom center), including contributions from particle-production hotspots, with the absolute value of the fractional hotspot contributions shown in the bottom panels.  The shaded region in each bottom panel represents the uncertainty from cosmic variance.  In each top panel we show the power spectra for $N_{{\mathrm{HS}}} = 0,1,10,100$ hotspots, as labeled (with $g=10$ and $\eta_*=150$~\text{Mpc}). Note that we have fixed fiducial values for our $\Lambda$CDM parameters and do not show variations in the power spectra arising from uncertainties on the parameters themselves. 
  }
  \label{fig:widefigure_powercors}
\end{figure*}
 
Using the harmonic coefficients from Equation~\eqref{eq:harmonicprofs}, the contribution to the $EE$ autospectrum takes a simple form:
\begin{align}
\label{eq.EE_hotspot}
    C^{EE}_{\ell,{\mathrm{HS}}}=\frac{1}{2\ell+1}\sum_{m=-\ell}^\ell\abs{t^E_{\ell m}}^2= \frac{\abs{\mathcal{I}^E_\ell}^2}{\pi^3}.
\end{align}
Equivalently, we can take $\mathcal{D}^{EE}_{\ell,{\mathrm{HS}}}\equiv \ell(\ell+1)/(2\pi)\,C_{\ell,{\mathrm{HS}}}^{EE} =\ell(\ell+1)/(2\pi^4)\,\abs{\mathcal{I}_\ell^E}^2$. An example of the hotspot contribution to the $TT, EE, \text{and}~TE$ power spectra is shown in Figure~\ref{fig:widefigure_powercors}, setting $\eta_* = 150~\text{Mpc}, \eta_{{\mathrm{HS}}}=\eta_{\rm rec}$, $g=10$, and $N_{{\mathrm{HS}}}=\{1,10,100\}$.  Note that the hotspot contribution to the power spectrum is quadratic in $f(k\eta_*)$, so it scales as\footnote{Here we consider hotspots a fixed comoving distance away (in this case $\chi_{\mathrm{HS}}=\chi_{\text{rec}}$), but distributed isotropically across the sky.} $g^2N_{{\mathrm{HS}}}$, where the $N_{{\mathrm{HS}}}$ factor enters because each hotspot is assumed to be uncorrelated, so we may add their contributions to the power spectrum linearly.

We find that the fractional contribution of the hotspots to the polarization power spectrum is comparable to that found for temperature.  
Additionally, as noted above, the contribution to the power spectrum goes as $g^2 N_{{\mathrm{HS}}}(g,M_0,\eta_*)$. For example, to get to $1\%$ of the fiducial $\Lambda$CDM power spectrum, one needs $g^2 N_{{\mathrm{HS}}} \simeq 10^4$. 

To proceed more rigorously, we can use Fisher forecast techniques to compute the $1\sigma$ sensitivity to $g^2 N_{{\mathrm{HS}}}$ from the power spectrum.  We can then find where the constraints become competitive with our localized search techniques. 

We consider the power spectrum of a CV-limited experiment (for which there is no noise contribution to the power spectrum), and treat $g^2 N_{\text{HS}}$ as the sole parameter in our forecast; note that $N_{\rm HS} = N_{\text{HS}}(g,M_0)$.  The Fisher matrix is thus one-dimensional: 
\begin{align}
    F_{g^2N_{\rm HS}} = & \sum_\ell \frac{(2\ell+1)}{2} \frac{f_{\rm{sky}}\left(\dv{C_{\ell,{\mathrm{HS}}}}{g^2N_{\rm HS}}\right)^2}{(C_\ell^{\Lambda\text{CDM}})^2} \\
    = & \sum_\ell \frac{(2\ell+1)}{2}\frac{f_{\rm{sky}} \left( C_{\ell,{\mathrm{HS}}}(g=1,N_{\rm HS}=1) \right)^2}{(C_\ell^{\Lambda\text{CDM}})^2}
\end{align}
The $1\sigma$ error on $g^2N_{\rm HS}$ is then given by $\sigma_{g^2N_{\rm HS}}=(F_{g^2N_{\rm HS}})^{-1/2}$. 

In practice, one would likely do a joint power spectrum analysis considering the joint $TT$/$EE$/$TE$ signal data vector, $\mathbf{d}(\ell)=(C_{\ell,{\mathrm{HS}}}^{TT},C_{\ell,{\mathrm{HS}}}^{EE},C_{\ell,{\mathrm{HS}}}^{TE})$.  The Gaussian covariance between these spectra is (with $X,Y \in [T,E]$)
\begin{align}
 \text{Cov}[C_{\ell}^{XY},C_{\ell}^{X'Y'}] = \frac{\Big(C_{\ell}^{XX'}C_{\ell}^{YY'}+C_{\ell}^{XY'}C_{\ell}^{YX'}\Big)}{(2\ell+1)f_\mathrm{sky}}  
 \label{eq:COVARIANCES}
\end{align}
The Fisher forecast for the joint $T$ and $E$ analysis is then 
\begin{align}
    F^{T\times E}_{g^2N_{\rm HS}} = & N_{{\mathrm{HS}}}^2 \sum_{\ell,X,X'}\mathbf{d}(\ell)\text{Cov}^{-1}\mathbf{d}(\ell) \nonumber \\ 
    = & N_{{\mathrm{HS}}}^2 \sum_{\ell,X,X'}C_{\ell,{\mathrm{HS}}}^X\text{Cov}^{-1}_{XX'}C_{\ell,{\mathrm{HS}}}^{X'} \,,
\label{eq:Fishjoint}
\end{align}
where we have taken $X=\{TT,EE,TE\}$ and we have noted that the derivative of the signal with respect to $g^2$ is simply given by the fiducial power spectra (evaluated at $g=1,N_{\text{HS}}=1$).

\begin{figure}[t!]
  \centering
  \includegraphics[width=0.48\textwidth]{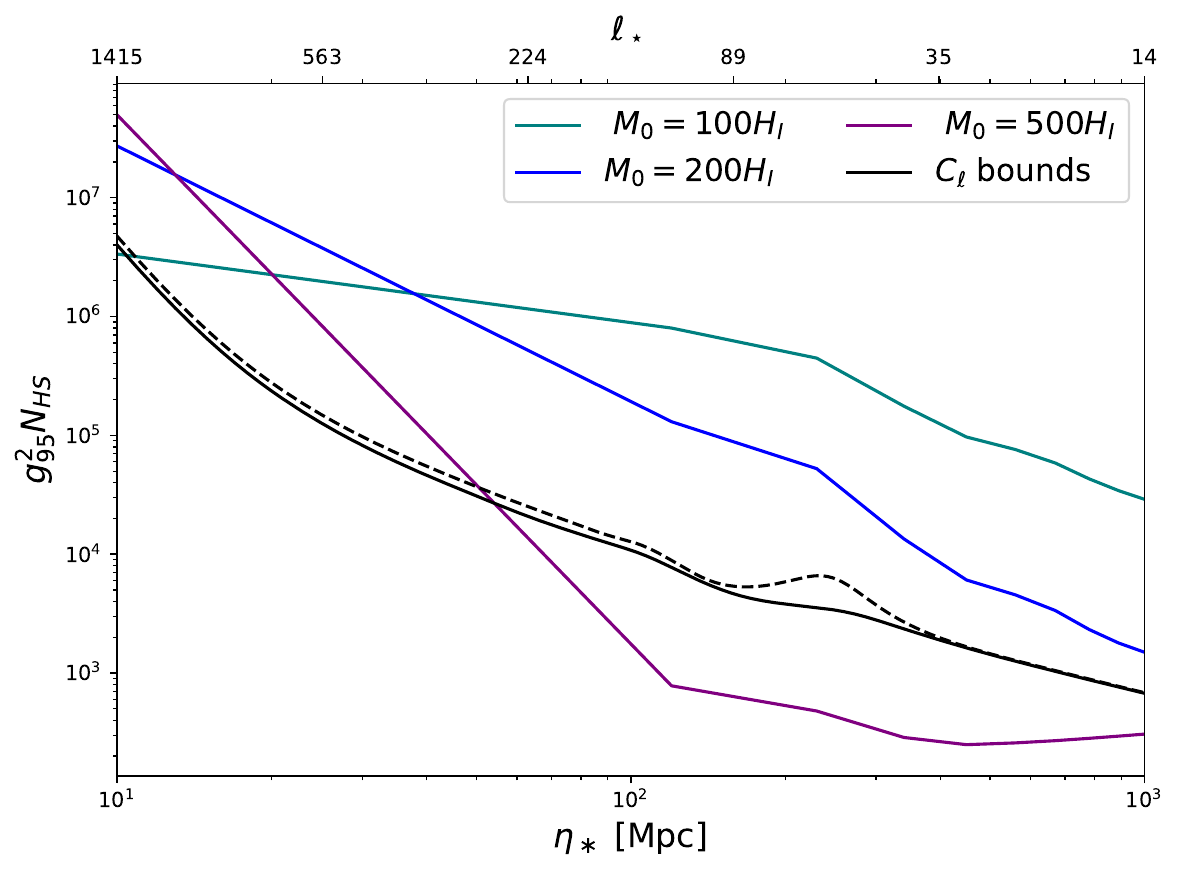}
  \caption{{Exclusion forecasts on $g^2N_{\rm HS}$ (at $95\%$ confidence) computed with Fisher matrix techniques for a CV-limited experiment with $f_{\text{sky}}=0.6$. The black lines denote constraints derived from the power spectrum (where solid uses $TT+EE+TE$ and dashed uses $EE$ only). The teal, blue, and purple lines show the $E$-mode matched-filter-derived constraints from Sec.~\ref{sec:bounds}, for $M_0/H_I = 100,200,500$, respectively. Notice that for lighter particles, where more are produced, the power spectrum constraints are stronger, but for more massive particles where fewer are produced, the matched filter approach provides significantly stronger constraints.}}
  \label{fig:widefigure_FISHforcs}
\end{figure}

In Figure~\ref{fig:widefigure_FISHforcs}, we compare our power-spectrum-based Fisher forecasts to the matched-filter-based exclusion curves computed for $M_{0} = \{100,200,500\}H_I$. This comparison is performed for a CV-limited experiment, and should be treated as a methodological study; we will return to a more careful treatment of power spectrum searches in future work. For larger-mass particles (\textit{e.g.,} $M_0=500H_I$), the matched-filter approach of this work provides tighter constraints than the power spectrum, with improvement factors around 10 for $\eta_* \geq 100 \,\,\text{Mpc}$. For lighter particles (which would be more numerous), the power spectrum bounds are somewhat better. This is as expected: in the limit of a large number of weak hotspots, the likelihood becomes asymptotically Gaussian, thus the power spectrum search is optimal.
  
We have also fixed to a specific fiducial $\Lambda$CDM cosmology here and ignored the variations in the power spectrum associated with shifts in the cosmological parameters. In a true power spectrum search these would be partially degenerate with an inflationary hotspot signal, and a full parameter analysis would be required, which would lead to further degradation of the constraints, particularly in cases where the hotspot contributions are nearly in phase with the $\Lambda$CDM power spectrum (as in the example in Figure~\ref{fig:widefigure_powercors}).\footnote{Note that the hotspot contributions are nearly in phase with the $\Lambda$CDM power spectrum for $\eta_* \approx r_s^{\rm rec}$, but we find that this behavior does not persist for much smaller or larger $\eta_*$ values.} This illustrates another advantage of the direct hotspot search, \textit{i.e.}, we are not limited by $\Lambda$CDM parameter degeneracies.

\section{Summary and Discussion}
\label{sec:summary}

\noindent 
Non-adiabatic production of massive particles is a feature appearing in many multi-field inflationary scenarios, and is known to leave observational signatures in the CMB. 
Ref.~\cite{Philcox:2024jpd} found that such signatures can be effectively constrained by applying a localized search to \emph{Planck} temperature data, but could easily be missed by standard correlation function analyses. In this work we expand the horizons of localized CMB searches from temperature to polarization, and build a new Poissonian likelihood, in analogy with likelihoods used in cluster abundance cosmology.  The use of our improved likelihood strengthens constraints on the coupling between the inflaton and an extremely massive particle by more than a factor of $\sim10$ for particles in the lower end of our probed mass range (\emph{e.g.}, $M_0=100H_I$), allowing us to probe perturbative (and thus self-consistent) regimes of the theory.

The use of polarization data has several advantages over temperature. Beyond providing an independent probe of inflationary physics, polarization comes with different noise properties, and because the Sunyaev-Zel'dovich effect is unpolarized to leading order there are many fewer possible false detections. Additionally, unlike the foregrounds, the hotspots should not have counterparts in $B$-mode maps, which gives us an additional method to verify any candidates. We also perform forecasting for current and future ground-based CMB experiments, which suggests that while the \emph{Planck} $T$ and $E$-mode constraints are comparable, in future experiments polarization will become a superior channel in which to look for such hotspots. Although this may be a model-dependent phenomenon, we expect that this is a relatively generic property of profile-finding searches for inflationary signatures given that it has long been known that polarization can constrain parameters better in the CV limit (e.g., \cite{Galli_2014}).

We validate our $E$-mode search pipeline by injecting hotspots into simulated component-separated data, and successfully recover the input parameters for sufficiently high values of $\eta_*$ and $g$. From the \emph{Planck} polarization data, we find no evidence of new physics, with no evidence for hotspots with SNR~$\geq6$ (we impose a high SNR threshold due to the four free parameters in the template used in the search).  This yields a relatively strong bound on couplings between the inflaton field, $\varphi$, and a much heavier field with $M_0/H_I \gtrsim 100$. For $M_0=100H_I$, we constrain the coupling $g\leq 2$ (95\% C.L.) for all values of $\eta_*$ considered here.  
 
Our analysis of \emph{Planck} $E$-mode polarization data is verified on both \textsc{sevem} and \textsc{smica} maps, which produce consistent results, with neither finding strong hotspot candidates. Our work improves substantially on the results of Ref.~\cite{Philcox:2024jpd}, and our polarization analysis puts strong bounds on massive scalars with mass $M_0 \sim 100H_I$. The inflationary Hubble scale is not known, but the current upper limit from primordial $B$-mode searches is $H_I<4.8\times 10^{13}~\text{GeV}$~\cite{BICEP:2021xfz}. This implies that our analysis may directly probe physics near the Grand Unified Theory scale, if $H_I$ is near the current upper limit. 

We also present a direct comparison of our profile-finding analysis method versus a power spectrum analysis. {We find that for more massive particles, localized searches analyzed with a Poissonian likelihood dominate over bounds inferred from the power spectrum, but in parameter regimes where many particles are produced the power spectrum (or other low-order correlators) will prove more constraining (this is natural in situations such as those studied in \cite{2020A&A...641A...9P,Chung_2000})}. We emphasize a generic point, which is that our analysis and forecasts point to the fact that for rare dramatic events local searches are preferable, where by contrast, for common events with smaller individual impacts, the best way forward is the measurement of low-order correlators. 

This work can be extended in several important ways. Minor improvements ($\sim10-20\%$) to the \textit{Planck} {matched filter sensitivity} could be obtained by jointly analyzing the \emph{Planck} PR4 temperature and $E$-mode polarization data. However, a much more exciting way forward is to apply our methods to higher-resolution CMB data.  Figure~\ref{fig:exclusions} shows that ACT is close to CV-limited on small and intermediate scales, and would yield a factor of improvement between $1.2-2.75$ for $E$ sensitivity, providing a clear motivation for a search using this data. Similarly, one could explore analyses using other high-resolution experiments, such as the South Pole Telescope~\cite{Bleem_2022,SPT-3G:2025bzu} or the Simons Observatory~\cite{SimonsObservatory:2018koc,SimonsObservatory:2025wwn}. Additionally, our pipeline can be simply extended to search for a wide variety of primordial features~\cite{M_nchmeyer_2019,Mirbabayi_2015}, and even for localized signatures induced by novel later-time physics such as patchy screening by axions~\cite{Mondino:2024rif,Goldstein:2024mfp} or dark photons~\cite{Pirvu:2023lch,McCarthy:2024ozh}.

\begin{acknowledgments}
We thank Soubhik Kumar for insightful discussions and comments on a draft of this work.  We also thank the anonymous referee for helpful comments.  LHA acknowledges support from the Columbia College Summer Research Fellowship, Columbia University, and the Barry Goldwater Scholarship. OHEP was a Junior Fellow of the Simons Society of Fellows. JCH acknowledges support from DOE grant DE-SC0011941, NASA grants 80NSSC22K0721 [ATP] and 80NSSC23K0463 [ADAP], the Sloan Foundation, and the Simons Foundation. This work utilized {\tt numpy}~\cite{harris2020array}, {\tt matplotlib}~\cite{Hunter:2007}, {\tt healpy}~\cite{Zonca2019}, and {\tt HEALPix}~\cite{2005ApJ...622..759G}. We acknowledge computing resources from Columbia University's Shared Research Computing Facility project, which is supported by NIH Research Facility Improvement Grant 1G20RR030893-01, and associated funds from the New York State Empire State Development, Division of Science Technology and Innovation (NYSTAR) Contract C090171, both awarded April 15, 2010. 
\end{acknowledgments}

\appendix

\section{\label{Appendix:appendix A} In-in computation of the curvature perturbation due to a massive particle}
\noindent Here we derive the curvature perturbation due to a massive particle.  Our treatment is very similar to that of~\cite{Kim:2021ida}; here our primary purpose is to be pedagogical.  We work in units with $c=1$, and the derivative with respect to cosmic time is represented by a dot.

Given the natural assumption that particle production happens when the heavy field's time-dependent mass reaches its minimum, the mass should rapidly increase after production, and we thus assume that the particle quickly becomes non-relativistic. This leads to the action: 
\begin{align}
    S_{\text{particle}}=-\int \dd t\sqrt{-g_{00}}M(t) \,,
\end{align}
where $g_{\mu\nu}$ is the metric. 
We parametrize the metric fluctuations in the standard ADM variables~\cite{Arnowitt_2008}, writing
\begin{align}
\dd s^2&=-N^2\dd t^2+h_{ij}(\dd x^i+N^i\dd t)(\dd x^j+N^j\dd t) \\
h_{ij}&=e^{2Ht}[(1-2\psi)\delta_{ij}+\gamma_{ij}]\nonumber \,,
\end{align}
where we choose $\gamma_{ij}$ to be transverse and traceless. In this gauge there are no inflaton fluctuations, $\delta\varphi=0$. It is proven in~\cite{Maldacena_2003} that $N$ and $N^i$ are time-independent at leading order. Furthermore, $N$ obeys the algebraic identity
\begin{align}
    N=1-\frac{\dot{\psi}}{H_I}=1+\frac{\dot{\zeta}}{H_I} \,,
\end{align}
where $\zeta$ is the comoving curvature perturbation:
\begin{align}
    \zeta=-\psi-H_I\frac{\delta\varphi}{\dot{\varphi}}=-\psi
\end{align}
in our gauge. We can then write our particle action more explicitly, 
\begin{align}
    S_{\text{particle}}=-\int \dd tNM(t)=-\int \dd t \left( M(t)+\dot{\zeta}\frac{M(t)}{H_I} \right) \,.
\end{align}
The second term induces the contributing curvature perturbation in our model. It will be more convenient to do our calculation in conformal time $\eta$ ($\dd \eta \equiv \dd t/a$), noting that we may write the relevant term in the action as
\begin{align}
    S_{\text{particle}} &\supset& &-\int \dd \eta \, \partial_\eta\zeta\frac{M(\eta)}{H_I}\\
    &=& &-\int_{\eta_*}^0\frac{M(\eta)}{H_I}\int \frac{\dd^3\bf{k}}{(2\pi)^3}\partial_\eta\zeta_{\vb{k}}e^{i\vb{k}\vdot \vb{x}_{{\mathrm{HS}}}}\nonumber
\end{align}
Also note that we can write the comoving curvature perturbation in terms of creation and annihilation operators, for a simple inflationary model~\cite{baumann2012tasilecturesinflation}:
\begin{align}
    \zeta_{\vb{k}}=\frac{H_I^2}{\dot{\varphi}_0\sqrt{2k^3}}[(1-ik\eta)e^{ik\eta}a^\dagger_{\vb{k}}+(1+ik\eta)e^{-ik\eta}a_{-\vb{k}}]
\end{align}

From here we have the requisite tools to compute the average curvature perturbation associated with the massive particle in the in-in formalism. Recall that in the in-in formalism we have the master formula in the commutator form in the interaction picture~\cite{Weinberg_2005,Lee:2024sks} for the expectation value of some operator, $\mathcal{O}(\eta)$, given some interaction Hamiltonian, $H_{\text{int}}$: 
\begin{align}
    & \langle \mathcal{O}(\eta)\rangle = \sum_{N=0}^{\infty}i^N\int_{-\infty}^{\eta} \dd \eta_N...\int \dd \eta_1 \\
    & \times\bra0\big[H_{\text{int}}(\eta_1),\big[H_{\text{int}}(\eta_2),...\big[H_{\text{int}}(\eta_N),\mathcal{O(\eta)}\big]...\big]\big]\ket0\nonumber
\end{align}
Note that to leading order this is simply
\begin{align}
    \langle\mathcal{O}(\eta)\rangle=i\int_{-\infty}^\eta \bra0\comm{H_{\text{int}}(\eta')}{\mathcal{O}(\eta)}\ket0\dd \eta' 
\end{align}
One can then observe that, by hermiticity, $\bra0\comm{H_{\text{int}}(\eta')}{\mathcal{O}(\eta)}\ket0=2i\Im{\langle H_{\text{int}}\mathcal{O}\rangle}$. From here we can note that
\begin{align}
    &\langle\zeta_{\vb{k}}(\eta\rightarrow0)\rangle\nonumber = \\
    &-2\lim_{\eta\rightarrow0}\int_{\eta_*}^0\dd \eta'\frac{M(\eta')}{H_I}
    \int \frac{\dd^3\vb{p}}{(2\pi)^3}e^{i\vb{p\vdot x_{{\mathrm{HS}}}}}\Im{\langle \partial_\eta\zeta_{\vb{p}}\zeta_{\vb{k}} \rangle}
\end{align}
Evaluating the central matrix element yields 
\begin{align}
   \lim_{\eta\rightarrow0} \langle \partial_\eta\zeta_{\vb{p}}\zeta_{\vb{k}}\rangle=\frac{H_I^4}{2\dot{\varphi}_0^2}\times\frac{p^2\eta'}{\sqrt{p^3k^3}}e^{-ip\eta'}\langle a_{-\vb{p}}a^\dagger_{\vb{k}} \rangle \nonumber\\
   =\frac{H_I^4}{2\dot{\varphi}_0^2}\times\frac{p^2\eta'}{\sqrt{p^3k^3}}e^{-ip\eta'}\times(2\pi)^3\delta^{(3)}(\vb{p}+\vb{k})
\end{align}
Taking the imaginary part and integrating over $\mathbf{p}$ gives us 
\begin{equation}
        \langle \zeta_{\vb{k}}\rangle=\frac{H_I^3}{\dot{\varphi_0}^2}\int_{\eta_*}^0\dd \eta'\frac{M(\eta')\eta'}{k}\sin{k\eta'}e^{-i\vb{k\vdot x_{{\mathrm{HS}}}}}
\end{equation}
This integral can be done if we approximate the mass around $\eta_*$ by dropping the constant $M_0$ term, such that $M\simeq\frac{g\dot{\varphi_0}}{H_I}\ln{\abs{\eta_*/\eta}}$. This then yields the hotspot profile from Sec.~\ref{sec:theory}: 
\begin{align}
     \langle \zeta_{\vb{k}}\rangle=\frac{gH_I^2}{\dot{\varphi_0}}e^{-i\vb{k\vdot x_{{\mathrm{HS}}}}}\frac{\text{Si}(k\eta_*)-\sin{(k\eta_*)}}{k^3} \,,
\end{align}
where $\text{Si}(x)=\int_0^x\frac{\sin{t}}{t}\dd t$ is the integrated sine function.

\section{Visual inspection of the candidates}\label{app:vis}

\begin{figure*}[t]
  \centering

  \begin{minipage}{0.96\textwidth}
    \centering
    \includegraphics[width=\textwidth]{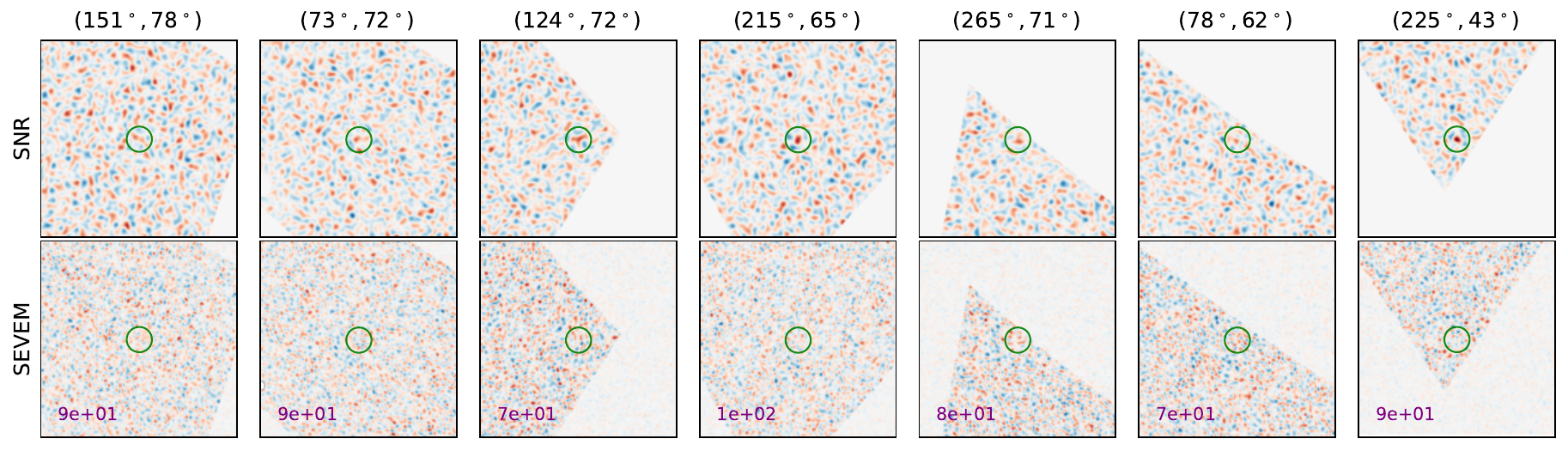}
    
  \end{minipage}
  
  \vspace{0.5cm}

  \begin{minipage}{0.96\textwidth}
    \centering
    \includegraphics[width=\textwidth]{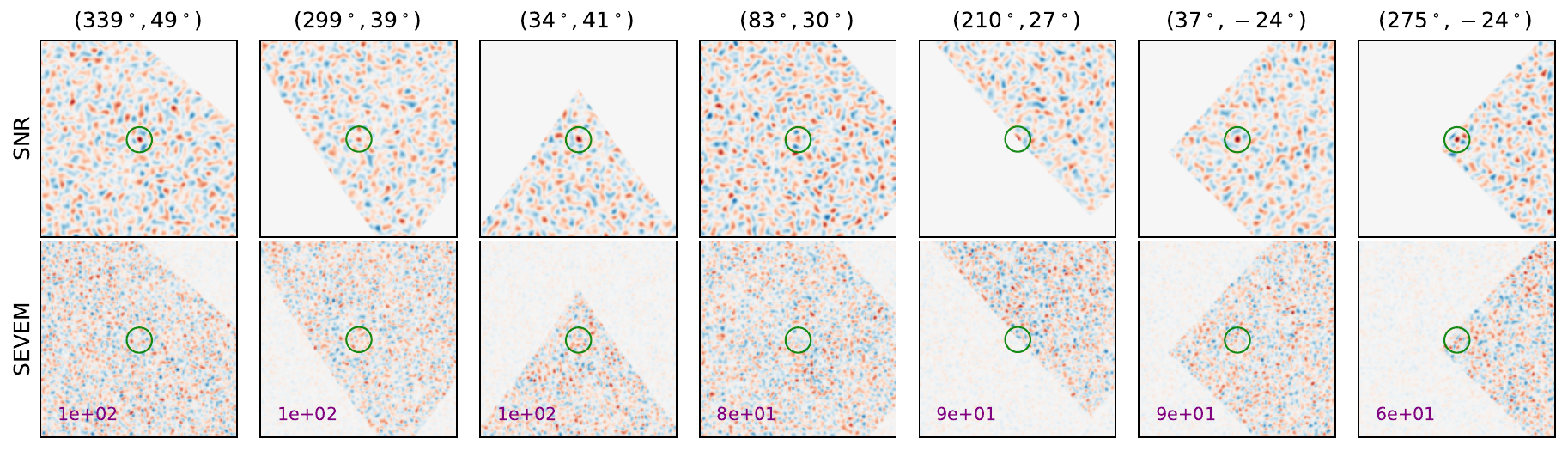}
    
  \end{minipage}

  \vspace{0.5cm}

  \begin{minipage}{0.96\textwidth}
    \centering
    \includegraphics[width=\textwidth]{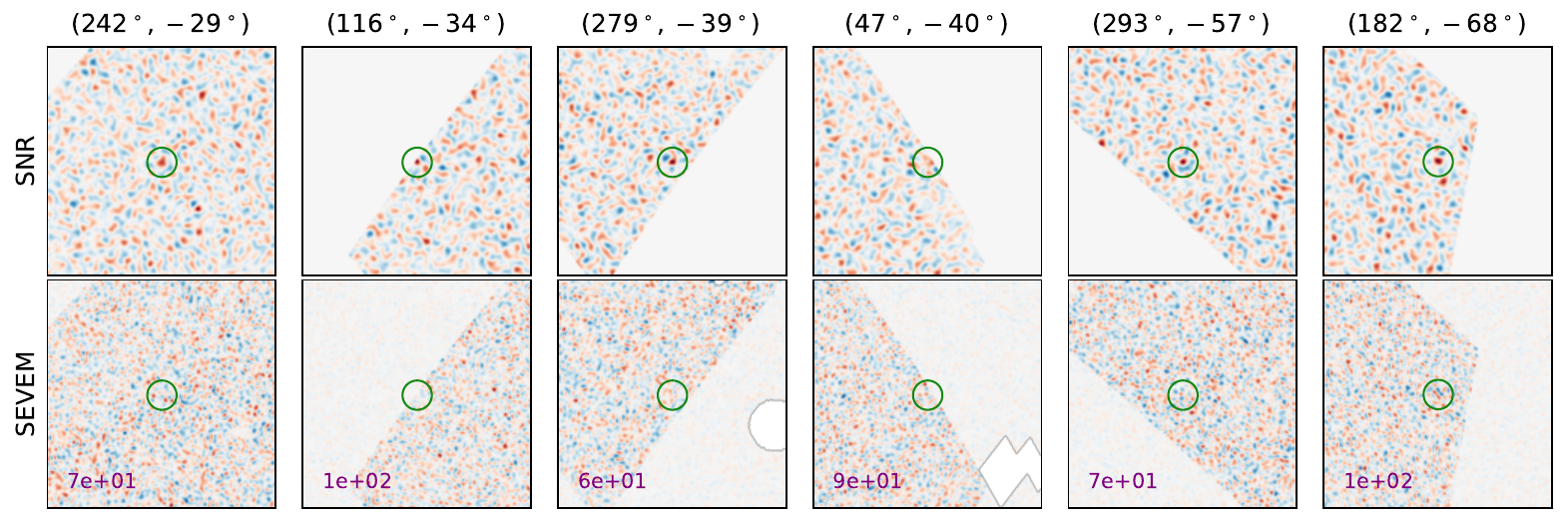}
    
  \end{minipage}

  \caption{SNR maps (top row) and \textsc{sevem} component-separated $E$-mode maps (second row) for all SNR $\geq 5$ hotspot candidates, which are circled in green. Each image is a $5^\circ\times5^\circ$ Cartesian projection around the centers given in the title. The SNR maps have colorbar truncated to $\pm5$, while the maximal absolute value for each other plot is shown in purple text. The bounding rectangles presented are merely the flat-space tiles used in our \textsc{szifi} analysis.}
  \label{fig:vertical_stacked}
\end{figure*}

\noindent In Figure~\ref{fig:vertical_stacked}, we present a visual inspection of the results of our polarization hotspot search.  We plot the \textsc{sevem} $E$-mode map at the locations of all of the SNR~$\geq5$ hotspot candidates. We also show SNR maps (\textit{i.e.}, ${\hat{g}}/{\sigma_g}$) on the top row.  We note that many candidates are found near the edges of the mask, and as such are likely ringing as opposed to primordial.  We also emphasize that due to the four parameters in our search template, moderately high SNR values are to be expected due to chance fluctuations.

\section{\textsc{SMICA} hotspot candidates}\label{app:smica}
\noindent We present the recovered parameters from the application of our analysis to the \emph{Planck} \textsc{smica} maps in Table~\ref{tab:smicacands}. We emphasize that beyond the fact that the \textsc{sevem} and \textsc{smica} maps produce a similar number of candidates, 20 versus 23, the candidates are also recovered with similar parameters. This demonstrates the robustness of our pipeline. 

\begin{table}[h!]
\centering
\begin{tabular}{cccccc}
\hline
   SNR &      Longitude [${}^\circ$] &      Latitude [${}^\circ$] &         $\hat{g}$ &   $\eta_*$ [Mpc] &   $\eta_{{\mathrm{HS}}}$ [Mpc] \\
\hline
 5.2 & 144.4   &  81.8 &   29  &    46.4 &  242.4 \\
 5.0 & 114.8  &  81.5 &   13   &   215.4  &  221.7 \\
 5.2  & 214.6  &  64.6 &   28  &    46.4 &  259.3 \\
 5.0 & 230.9 &  63.6 &    10 &   599.5  &  240.0 \\
 5.1 & 265.5  &  71.2 &   12  &   129.2  &  292.2 \\
 5.2 &  78.4 &  62.4 &   15  &   215.4  &  378.4 \\
 5.3 & 339.2  &  59.3  &   10  &   359.4  &  290.8 \\
 5.1 & 224.6  &  43.1 &  293   &    16.7  &  294.1 \\
 5.0 & 106.9  &  36.1  &   46  &    27.8 &  257.7 \\
 5.0 & 275.4   &  40.4 &   71  &    27.8 &  257.7 \\
 5.1 &  74.3 &  33.6 &  902   &    10      &  272.2 \\
 5.0 &  82.8   &  30.5 &   24  &    46.4 &  242.4 \\
 5.0 &  93.1 &  25.7 &   19  &    77.4 &  343.8 \\
 5.5 & 324.0  &  21.6 &  106   &    27.8&  293.1 \\
 5.1 &  78.5 & -34.5 &   28  &    46.4 &  242.4 \\
 5.4  & 116.2  & -33.7 & 1194    &    10      &  272.2 \\
 5.1 & 270.8  & -45.6 &  186   &    16.7  &  266.8 \\
 5.3  & 342.3  & -41.0 &   19  &    77.4 &  231.2 \\
 5.1 & 356.5  & -40.7 & 1052     &    10      &  272.2 \\
 5.1 &  29.1 & -46.1 & 1094    &    10      &  272.2 \\
 5.0 & 121.3  & -44.3 &   91  &    27.8 &  303.2\\
 5.3 & 292.7  & -57.1 &  859  &    10      &  272.2 \\
 5.2 & 181.8  & -67.5 &   87  &    27.8 &  288.0 \\
\hline
\end{tabular}
\caption{Inferred parameters for the hotspot candidates from \emph{Planck} \textsc{smica} $E$-mode maps. Notice that none of them are strong candidates (SNR~$\geq6$), and that almost all have direct counterparts with candidates from the \textsc{sevem} maps.} 
\label{tab:smicacands}
\end{table}

\section{\label{Appendix:appendix D} Derivation of the analytic selection function and comparison with simulations}
\noindent Here we derive the selection function from the variance intrinsic to our matched-filter estimator. We can see from our construction in Eq.~\eqref{eq:MMF} that we may treat $\hat{g}$ as a random variable characterized by a Gaussian distribution with standard deviation $\sigma_g$. Given a true hotspot at fixed angular position with a true value for the coupling given by $g$, we expect the probability distribution of the estimated $\hat{g}$ value from our matched filter to be given by 
\begin{align}
    \mathbb{P}[\hat{g}]=\frac{1}{\sqrt{2\pi \sigma_g^2}}\exp{-\frac{(\hat{g}-g)^2}{2\sigma_g^2}}~.
\end{align}
We identify hotspot detections as those locations where $\hat{g}\geq 5\sigma_g$.  We then seek to compute the selection function, \textit{i.e.,} the fraction of isotropically distributed hotspots that we expect our matched filter to detect. We can integrate the above probability distribution to obtain this fraction. Let us first define an auxiliary selection function $S(g,\eta_*,\eta_{\text{HS}})$, which is not yet integrated over the sphere:
\begin{align}
    S(g,\eta_*,\eta_{\text{HS}})&=\frac{1}{\sqrt{2\pi\sigma_g^2}}\int_{5\sigma_g}^{\infty} \dd \hat{g} \exp{-\frac{(\hat{g}-g)^2}{2\sigma_g^2}}\nonumber \\&=\frac{1}{\sqrt{2\pi}}\int_{5-\frac{g}{\sigma_g}}^{\infty} e^{-x^2/2}\dd x \,.
\end{align}
Note that the right-hand side depends on $\eta_*$ and $\eta_{\rm HS}$ via their influence on $\sigma_g$ (cf.~Eq.~\eqref{eq.var}).  The integral can be evaluated in terms of the complementary error function $\text{erfc}(z)\equiv\frac{2}{\sqrt{\pi}}\int_z^{\infty} e^{-t^2}\dd t$, yielding
$$
S(g,\eta_*,\eta_{\rm HS}) = \frac{1}{2}\text{erfc}\left(\frac{1}{\sqrt{2}}\left(5-\frac{g}{\sigma_g}\right)\right) \,.
$$
We then formally average this result over the whole sky, $\mathcal{S} = 1/(4\pi) \int S \dd^2 \hat{n} \simeq f_{\text{sky}}S$. Here we have assumed that the noise is close to isotropic across the sky. Written explicitly, we thus find
\begin{align}
    \mathcal{S}(g,\eta_*,\eta_{\rm HS}) = f_{\text{sky}}\frac{1}{2}\text{erfc}\left(\frac{1}{\sqrt{2}}\left(5-\frac{g}{\sigma_g}\right)\right) \,, 
\end{align}
as in Eq.~\eqref{eq:selection_function}.  Note that this result is consistent with the tSZ cluster selection function computed in Ref.~\citep{Planck:2015vgm} upon appropriate renaming of variables.

\begin{figure}[t]
  \centering
  \includegraphics[width=0.48\textwidth]{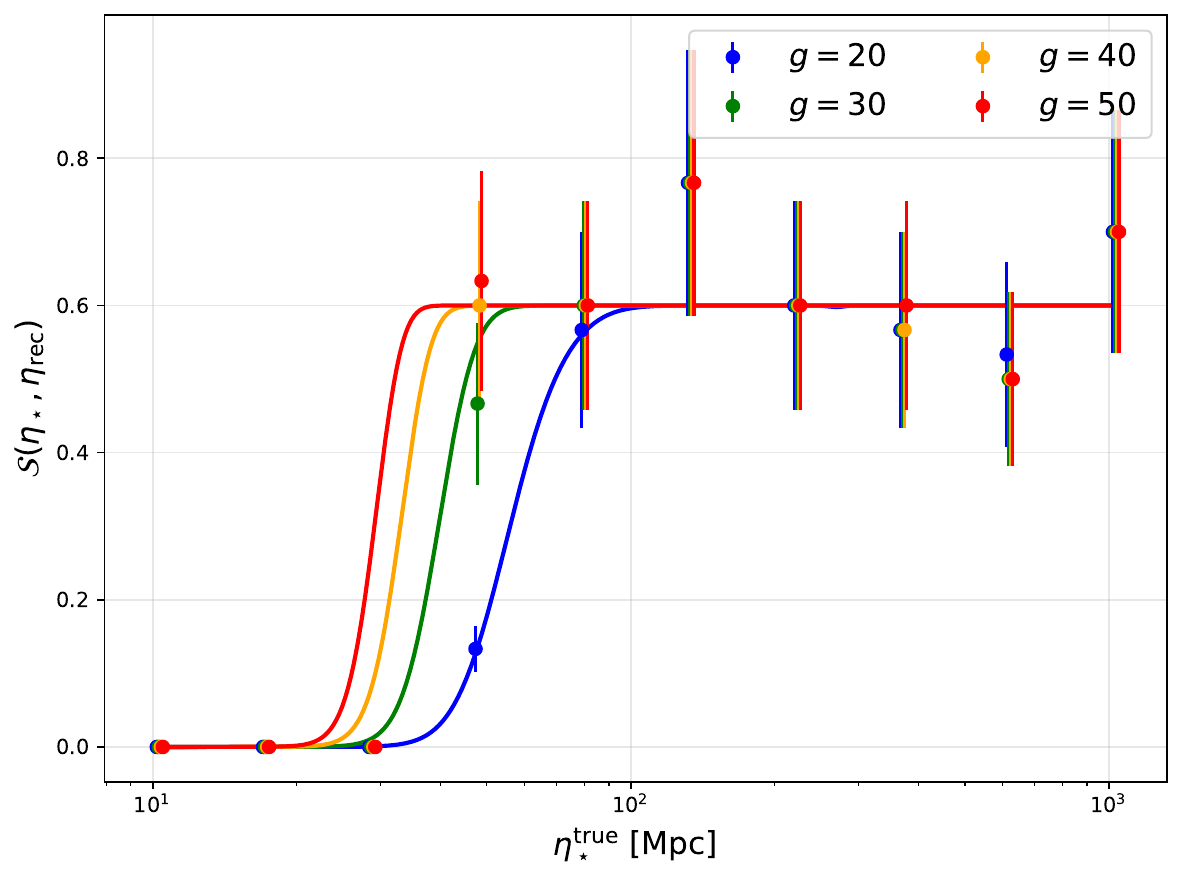}
  \caption{Here we demonstrate agreement between the analytically computed selection function in Eq.~\eqref{eq:selection_function}, shown in the solid curves, and the selection function estimated from numerical simulations, shown in the data points with error bars.  For simplicity, here we simulate hotspots on the surface of last scattering ($\eta_{\rm HS} = \eta_{\rm rec}$). We find sufficiently close agreement to use the analytic selection function in our likelihood analysis.}
  \label{fig:selectionplot}
\end{figure}

It is also worth directly comparing this analytic result to the simulation-derived selection function shown in the central panel of Fig.~\ref{fig:widefigure_sims}. Note that since we only inject 30 hotspots for each $\eta_*$ value, we expect relatively large uncertainties associated with the simulation-based measurement of the selection function. Assuming approximately Poissonian errors ($\frac{1}{\sqrt{f_{\rm{sky}}\times 30}}\simeq 23.5\%$), we see in Fig.~\ref{fig:selectionplot} that the analytic and simulation-derived results are consistent within $1\sigma$. Note also that because our pipeline injects 30 simulated hotspots at each given $\eta_*$, then rescales the hotspot profiles for each value of $g$, the fluctuations in the simulation-based estimates are highly correlated across $g$ values, as is evident in Fig.~\ref{fig:selectionplot}. In particular, the fluctuations above the maximum of the selection function would vanish in the limit of an infinite number of simulations (if the injected hotspots are truly distributed isotropically across the sky it would be impossible to have $\mathcal{S} > f_{\rm{sky}}$). The consistency of the analytic and simulation-based results within the $1\sigma$ error bars is satisfactory for our purposes, and massively reduces the required computational expense.  In Sec.~\ref{sec:sims}, we note that for each value of $g$ and $\eta_{\rm{HS}}$ (scanning over $\eta_*$), the simulations required roughly 20 CPU-hours; as such, with an analytical form of the selection function the full likelihood procedure becomes much more computationally tractable.

\begin{figure}[t]
  \centering
  \includegraphics[width=0.48\textwidth]{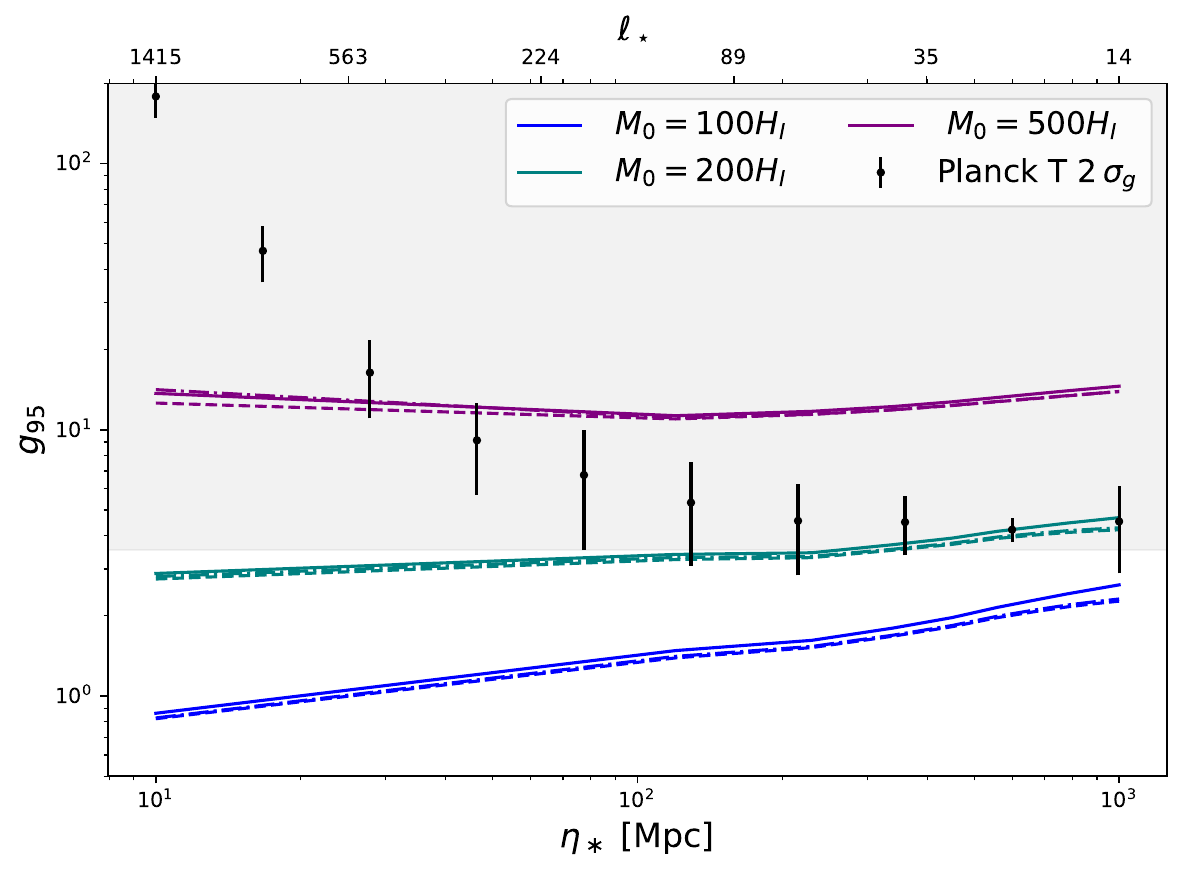}
  \caption{Exclusion bounds on $g$ derived from the \emph{Planck} CMB temperature data (dash-dotted), updating the results from Ref.~\cite{Philcox:2024jpd}.  We also show forecast bounds derived from ACT (solid) and a CV-limited experiment (dashed). For sufficiently light particles we find that CMB temperature data can constrain perturbatively small values of $g$.  For comparison, we also show the sensitivity-based $95\%$ confidence upper limits derived in Ref.~\cite{Philcox:2024jpd} (black points), which are supplanted by the new exclusion bounds presented here. For lower-mass particles and/or lower $\eta_*$ values, which lead to large numbers of particles being produced, the Poissonian-likelihood-derived constraints are much tighter than the single-hotspot sensitivities. As in Fig.~\ref{fig:trueexclusions}, we denote the nonperturbative region in $g$ by the gray shaded region.}
  \label{fig:TEMPconstraints}
\end{figure}

\section{\label{Appendix:appendix E} Update to previous temperature-based constraints }

\noindent Ref.~\cite{Philcox:2024jpd} presented constraints on inflationary particle production using a similar matched-filter approach to that considered here, but using only CMB temperature data.  However, they did not build a full Poissonian likelihood for the signal; instead, they only considered the single-hotspot $\sigma_g$ sensitivity, akin to that computed in Sec.~\ref{sec:forecasts}.  Here, we revisit and correct this oversight, providing full likelihood-based constraints from CMB temperature data.

Given that no temperature hotspots were detected in Ref.~\cite{Philcox:2024jpd}, the only update to our likelihood in Eq.~\eqref{eq:likelihood_simple} that is needed to perform this computation is to change the transfer function and noise from the $E$-mode field to the $T$ field.  These changes modify the selection function, which we represent by changing it from a function of $\sigma_g^E$ to $\sigma_g^T$.\footnote{One could also make a further extension to consider the Poissonian likelihood for a joint matched-filter search in $T$ and $E$; we do not consider this here.}  Here, we compare the $T$-based bounds derived from this likelihood to the sensitivities from Ref.~\cite{Philcox:2024jpd}, and include specific results for temperature constraints that were not the focus of Sec.~\ref{sec:bounds}, but are important results of this updated analysis. The $g$ exclusion bounds are shown in Fig.~\ref{fig:TEMPconstraints}. For particles in the lower-mass regime of our search, we obtain significant improvements in the bounds. This is particularly evident in the low $\eta_*$-regime (describing particle production later during inflation): the full likelihood constraints on $g$ improve by up to a factor of $\sim10$ (for $M_0=500H_I$) to $\sim100$ (for $M_0=100H_I$). As the expected number of particles decreases, the constraints weaken; if $N_{\rm{pred}}<1$, the full likelihood should yield weaker constraints than expected from the single-hotspot sensitivity approach (which ignores the abundance information). As such, we find that Ref.~\cite{Philcox:2024jpd} obtained overly optimistic constraints on very massive particles ($M_0\geq 500H_I$), except at low $\eta_*$. We also present forecast constraints for ACT and a CV-limited experiment. 
Interestingly, though not surprisingly, on these scales we do not expect to see much improvement from current and future $T$ data, which is to say that on the relevant corresponding $\ell$ scales, \emph{Planck} is CV-limited. In sum, the full likelihood approach significantly improves the \emph{Planck}-derived bounds on $g$ for lower-mass particles, and as in $E$ allows us to push into the perturbative regime of the coupling.

\bibliography{Hotspot_paper}

\end{document}